\documentclass{article}
\usepackage{amsmath}
\usepackage{amssymb}
\usepackage{amsfonts}
\usepackage{mathrsfs}
\usepackage{makeidx}
\usepackage{tikz}
\usepackage{dsfont}
\usepackage{tikz}
\usepackage{amsthm}
\usepackage{physics}
\usepackage{tensor}
\usepackage[affil-it]{authblk}
\usepackage{changes}
\usepackage{pdfcomment}

\usepackage[margin=1.3in]{geometry}

\usetikzlibrary{decorations.pathmorphing}
\tikzset{zigzag/.style={decorate, decoration=zigzag}}

\renewcommand{\restriction}{\mathord{\upharpoonright}}
\newcommand*\diff{\mathrm{d}}

\newtheorem{theorem}{Theorem}[section]

\newtheorem{definition}[theorem]{Definition}

\begin{document}
	\title{Hadamard states on spherically symmetric characteristic surfaces, the semi-classical Einstein equations and the Hawking effect}
	\author{Daan W. Janssen%
		\thanks{Electronic address: \texttt{janssen@itp.uni-leipzig.de}}\, and Rainer Verch\thanks{Electronic address: \texttt{rainer.verch@uni-leipzig.de}}}
	\affil{Institute for Theoretical Physics,\\University of Leipzig,\\Germany}
	\maketitle
	\noindent We investigate quasi-free Hadamard states defined via characteristic initial data on null cones centred at the axis of symmetry in spherically symmetric space-times. We characterize the necessary singular behaviour of null boundary two-point functions such that one can define non-linear observables at this null boundary and give formulas for the calculation of these observables. These results extend earlier characterizations of null boundary states defining Hadamard states in the bulk of the null cone. As an application of our derived formulas, we consider their implications for the semi-classical Einstein equations and calculate the vacuum polarization associated with Hawking radiation near a collapsing body.
	
	\section{Introduction}
	For many purposes, it can be argued that Hadamard states form the preferred class of physical states on a linear scalar quantum field theory, see for instance Ref.\ \cite{fewsterArtState2018}. In particular, it is for this class of states that the expectation value of non-linear observables, such as the (renormalized) stress-energy tensor can be evaluated. A Hadamard state can be characterized by the singular behaviour of its $n$-point functions. As for a quasi-free Hadamard state it is in particular the two-point function that specifies the state on the entire algebra of linear observables, it is also the singular behaviour of this two-point function, as a bi-distribution on a space-time, that captures the Hadamard property. This singular behaviour can be characterized in multiple ways, most notably via an explicit asymptotic expansion of the integral kernel associated with the two-point function, as given in Ref.\ \cite{kayTheoremsUniquenessThermal1991}, or equivalently via the microlocal spectrum condition, as shown in Ref.\ \cite{radzikowskiMicrolocalApproachHadamard1996}. That these definitions are equivalent, is most neatly shown if we do not only consider the class of two-point functions of Hadamard states, but the slightly more general notion of a Hadamard paramatrices. In section \ref{sec:char_had} we will recall the precise characterisation of a Hadamard parametrix in terms of its asymptotic expansion as well as recalling the microlocal spectrum condition and the role of the commutation relation and equation of motion in this definition.
	
	On space-times that admit some characteristic Cauchy surface as a boundary, Hadamard states can be constructed using a bulk to boundary correspondence between the linear observables in the domain of dependence of this boundary surface and an algebra of observables that can be defined on this surface, as used in Ref.\ \cite{gerardConstructionHadamardStates2016}. In asymptotically flat space-times these ideas can be extended to conformal boundaries, in particular the past or future null boundary of this space-time, as shown in Ref.\ \cite{dappiaggiHadamardStatesLightlike2017}. As we will recall in more detail in Sec.\ \ref{sec:qftsetup}, on these boundary states one can consider an analogue of the microlocal spectrum condition that guarantees the induced state in the bulk to be Hadamard. This allows one to define expectation values for this state of locally covariant non-linear observables in the bulk, such as Wick squares and time-ordered products as defined in Ref.\ \cite{hollandsLocalWickPolynomials2001} and \cite{hollandsExistenceLocalCovariant2002}. However, generally such expectation values diverge near the boundary and hence this state cannot be extended as a Hadamard state across this null boundary. We shall argue that this is related to the fact that, unlike the situation for bulk Hadamard parametrices, the wave front set condition formulated in Ref.\ \cite{gerardConstructionHadamardStates2016} does not uniquely fix the singular behaviour of boundary two-point functions. We therefore introduce a necessary condition for extendibility of the bulk Hadamard state in section \ref{sec:char_had} to explicitly derive the singular part of the boundary two-point function.
	
	For the resulting class of boundary states, we can now show that non-linear observables near the boundary indeed remain finite. In Section \ref{sec:renorm} we derive explicit formulas for various non-linear observables at the boundary given a boundary two-point function. We shall show how these formulas are particularly useful in the context of semi-classical gravity if one wishes to study the semi-classical Einstein equations as a characteristic initial value problem.
	
	As a further application, in Sec.\ \ref{sec:hawking_rad} we shall also give a computation a Wick square expectation value of a quantum field near a gravitationally collapsing body. We shall also comment on how these methods can be generalized to calculate the contribution of the `Hawking radiation' produced in gravitational collapse to components of the (expectation value of the) stress-energy tensor near the collapsing body. These calculations may open the way to a new understanding of the Hawking effect and black hole evaporation in the context of semi-classical gravity.

	\section{Basic ingredients from geometry and field theory}
	For simplicity we will for now restrict our attention to spherically symmetric set-ups. Further generalizations we deem a priori possible, albeit calculationally heavy.
	\subsection{Geometrical set-up}
	\label{sec:geom}
	We consider a space-time $M$ (assumed to be globally hyperbolic) that is spherically symmetric and admits global coordinates $(t,r,\Omega):M\rightarrow \mathbb{R}\times\mathbb{R}_{\geq0}\times S^2$ (where this map defines a chart almost everywhere, in particular with the exception of the axis of symmetry $r=0$). We assume that in these coordinates the metric is of the form
	\begin{equation}
	\label{eq:metric}
	\diff s^2=-\exp(2(\alpha+\beta))\diff t^2+\exp(2\beta)(-2\diff r\diff t+r^2\diff \Omega^2),
	\end{equation}
	where $\alpha,\beta$ are functions of $t,r$. Note that in the limiting case that $\alpha=0$, this space-time is conformally flat, hence we shall refer to this coordinate system as \textup{quasi-conformal coordinates}. We should also point out that, as can be read off from for instance the formula for the Ricci scalar $\mathfrak{R}$, that not all smooth functions $\alpha,\beta$ are compatible with a space-time that is smooth at $r=0$. These regularity conditions shall be addressed when they become relevant for computations.\footnote{Admittedly this coordinate system differs from standard choices made to parameterize spherically symmetric space-times. In more conventional choices the coordinate $r$ is either chosen to be an affine parameter for null geodesics (see for instance Ref.\ \cite{hollandsstefanAspectsQuantumField2000}) or is chosen such that the volume of a two-sphere at $r=R$ has surface area $4\pi R^2$ (see Ref.\ \cite{christodoulouProblemSelfgravitatingScalar1986}). In the latter case, where the metric has the form 			 	   
	\begin{equation}
		\diff s^2=-A^2\diff t^2-B^2\diff R \diff t+R^2\diff\Omega^2,
	\end{equation} a coordinate transformation that brings the metric into the form of eq.\ \eqref{eq:metric} (assuming asymptotic flatness for simplicity) is given by \begin{equation}
	r(t,R)=\left(\int_R^\infty\diff R\frac{B^2}{R^2}\right)^{-1}.
	\end{equation}}
	
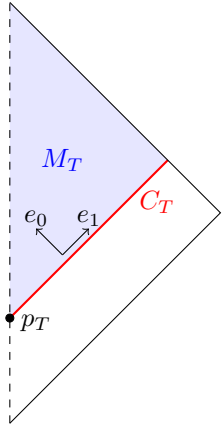
\begin{figure}[h]
		\centering
		\begin{tikzpicture}[scale=.7]
		\fill[blue!10] (0,2)--(3,5)--(0,8)--(0,2);
		\draw[blue] (1,5) node {$M_T$};
		\draw[->] (1,3.2)--(1.5,3.7);
		\draw[->] (1,3.2)--(.5,3.7);
		\draw (1.5,3.9) node {$e_1$} (.5,3.9) node {$e_0$};
		\draw[thick, red] (0,2)--(3,5);
		\draw[fill=black] (0,2) circle (.5ex) (0.5,1.9) node {$p_T$};
		\draw (0,0)--(4,4)--(0,8);
		\draw[dashed] (0,0)--(0,8);
		\draw[red] (2.8,4.2) node {$C_T$};
		\end{tikzpicture}
		\caption{A Penrose diagram sketching the geometric set-up}
		\label{fig:geo_setup}
	\end{figure}

	We refer to the point with $t=T, r=0$ as $p_T\in M$ and the hypersurface $t=T,r\neq0$ as $C_T$. The globally hyperbolic space-time $t>T$ will be denoted as $M_T:=I^+(C_T)$. Following Ref.\ \cite{christodoulouProblemSelfgravitatingScalar1986} we introduce a local null tetrad $(e_i)_{i=0,...,3}$, defined away from $r=0$ by
	\begin{equation}
		e_0=\partial_t-\frac{1}{2}\exp(2\alpha)\partial_r,\;e_1=\exp(-2\beta)\partial_r, 
	\end{equation}
	and $(e_2,e_3)=(r\exp(\beta))^{-1}(\zeta_1,\zeta_2)$ , where $(\zeta_1,\zeta_2)$ form a local orthonormal frame of $T_\Omega S^2$ (with standard metric $\diff\Omega^2$). Using this frame we can write
	\begin{equation}
		g^{\mu\nu}=\eta^{ab}e_a^\mu e_b^\nu,
	\end{equation}
	where $\eta_{ab}$ is the Minkowski metric expressed in double null coordinates, i.e.
	\begin{equation}
		\eta^{ab}=\eta_{ab}=\begin{pmatrix}
		0&-1&0&0\\
		-1&0&0&0\\
		0&0&1&0\\
		0&0&0&1
		\end{pmatrix}.
	\end{equation}
	This set-up is sketched in Fig.\ \ref{fig:geo_setup}.
	
	We can see that $C_T$ is a characteristic (i.e. null) hypersurface ruled by the bicharacteristics generated by $e_1$, i.e.  the null geodesics $\gamma_{T,\Omega}:(0,\infty)\rightarrow C_T$ satisfying 
	\begin{equation}
		\gamma_{T,\Omega}(\lambda)=(T,r(\lambda),\Omega),\;\dv{r}{\lambda}=\exp(-2\beta(T,r(\lambda))),\;\lim_{\lambda\downarrow0}r(\lambda)=0.
	\end{equation}
	Note in particular that $e_1$ is both tangent and orthogonal to $C_T$. However, as $e_1$ is a null vector, it can not be normalized. Furthermore, as the induced metric on $C_T$ is degenerate, one cannot use it directly to define a sensible $\mathbb{R}$-valued hypersurface element on $C_T$ to be used as an integration measure. Nevertheless, the vector hypersurface element, that for a spacelike smooth surface $\Sigma$ is defined as 
	\begin{equation}
		\diff \mathbf{S}=\diff x \diff y \diff z\sqrt{h}\mathbf{n},
	\end{equation}
	with $\mathbf{n}$ the future directed normal vector to $\Sigma$ and $h$ the determinant of the induced metric on $\Sigma$ (in coordinates $(x,y,z)$), can be generalized to null hypersurfaces as $C_T$. Using for instance a limiting procedure where $C_T$ is approached by space-like hypersurfaces, one finds that in our quasi-conformal coordinates, this vector hypersurface element takes the form
	\begin{equation}
		\diff \mathbf{S}=\diff r \diff \Omega r^2\exp(2\beta)\partial_r.
	\end{equation}
	As we will see in Sec.\ \ref{sec:qftsetup}, this vector valued integral measure naturally appears in the Klein-Gordon inner product, or alternatively the symplectic structure, for solutions to the Klein Gordon equation on $M_T$.
	
	\subsection{The classical equation of motion for the scalar field}
	On the space-times introduced above, we consider a real scalar quantum field theory. In particular, we consider a quantization of the real scalar field theory given by the equation of motion 
	\begin{equation}
	\label{eq:kg}
	(\Box-m^2-\xi \mathfrak{R})\phi=:P\phi=0,
	\end{equation}
	where classically we typically take $\phi\in\mathcal{E}(M)$ as a smooth function.\footnote{After quantization, the quantum field $\Phi$ can be seen to live in the space of `operator valued distributions' (or more generally, distributions taking values in a *-algebra) instead of smooth functions.} Here $m$ is some mass parameter, $\xi$ a dimensionless coupling, $\mathfrak{R}$ the Ricci curvature scalar and $\Box$ the Laplace-Beltrami operator given by 
	\begin{equation}
		\Box\phi=\nabla_\mu\nabla^\mu\phi=\frac{1}{\sqrt{\vert g\vert}}\partial_\mu(\sqrt{\vert g\vert}g^{\mu\nu}\partial_\nu\phi),
	\end{equation} with $\nabla$ the Levi-Civita connection. In our coordinate system (and using the vectorfields $e_0,e_1$ as derivations), we can rewrite the sourced equation of motion $P\phi=f\in\mathcal{D}(M)$ as
	\begin{equation}
		2(e_1\circ e_0)(r\exp(\beta)\phi)=\left((r\exp(\beta))^{-2}\Delta_\Omega-\tilde{V}\right)r\exp(\beta)\phi-r\exp(\beta)f,
	\end{equation}
	where $\tilde{V}$ is a modified potential and $(e_1\circ e_0)f=e_1^\mu\partial_\mu \left(e_0^\nu\partial_\nu f\right)$.
	\begin{equation}
		\tilde{V}=m^2+\xi\mathfrak{R}-2(r\exp(\beta))^{-1}(e_1\circ e_0)r\exp(\beta).
	\end{equation}
	Note that this additional term to the modified potential is reminiscent of the Ricci scalar, which equals
	\begin{align}
	\mathfrak{R}=&12(r\exp(\beta))^{-1}(e_1\circ e_0)r\exp(\beta)\nonumber\\&-2\exp(2(\alpha-\beta))\left(\frac{1-\exp(-2\alpha)}{r^2}-\frac{2\partial_r\alpha}{r}+2(\partial_r\alpha)^2+\partial_r^2\alpha\right).
	\end{align}
	In fact we can write
	\begin{align}
	\label{eq:vtilde}
	\tilde{V}=&m^2+\left(\xi-\frac{1}{6}\right)\mathfrak{R}\nonumber\\
	&-\frac{1}{3}\exp(2(\alpha-\beta))\left(\frac{1-\exp(-2\alpha)}{r^2}-\frac{2\partial_r\alpha}{r}+2(\partial_r\alpha)^2+\partial_r^2\alpha\right),
	\end{align}
	from which we can see that for a conformally coupled field, i.e. with $\xi=\frac{1}{6}$, on a conformally flat background, i.e. where $\alpha=0$, the modified potential takes the simple form $\tilde{V}=m^2$. Due to frequent appearance in the rest of this paper, it is convenient to define
	\begin{equation}\label{eq:R}
	R(t,r):=r\exp(\beta(t,r)).
	\end{equation}
	Defining the differential operator on $\mathcal{E}(C_T)$
	\begin{equation}
	K_T=\frac{1}{2}\left[\exp(2\beta(T,r))\tilde{V}(T,r)-\frac{1}{r^2}\Delta_\Omega-\partial_r\exp(2\alpha(T,r))\partial_r\right],
	\end{equation}
	we can shortly write the sourced equation of motion as
	\begin{equation}
	\label{eq:kgchar}
	(\partial_r\partial_t+K_t)R\phi=-\frac{R}{2}f.
	\end{equation}
	It is this form of the (sourced) Klein-Gordon equation that we shall use in most of the following considerations.\footnote{We shall often drop the subscript $t$ from $K_t$ when it is clear from context at what surface $C_t$ it is meant to operate.}
	
	\subsection{Quantum field theoretic set-up}
	\label{sec:qftsetup}
	As alluded to in the introduction, we can consider a quantum field theory to live equivalently on the bulk $M_T$ or on the boundary $C_T$. We shall quickly introduce both these viewpoints.
	\subsubsection{The bulk theory}
	\label{sec:qft_bulk_setup}
	We use the framework of algebraic quantum field theory, in particular using the model for the free real scalar quantum field due to Ref.\ \cite{dimockAlgebrasLocalObservables1980} in the formulation used for instance in Ref.\ \cite{beniniModelsFreeQuantum2015}. In order to fix conventions and notation, we shall quickly sketch the definition of the algebra of linear observables $\mathcal{A}(M)$ on the space-time $M$. This is a unital *-algebra consisting of polynomials (with complex valued coefficients) over objects of the form $\Phi(f)$ with $f\in\mathcal{D}(M)$ a (real-valued) test function on $M$ such that\footnote{One can work equally well with complex test functions, where for $h=f_1+\textup{i}f_2\in\mathcal{D}(M,\mathbb{C})$ we define \begin{equation}
		\Phi(h):=\Phi(f_1)+\textup{i}\Phi(f_2).
		\end{equation}
		This means in particular that
		\begin{equation}
			\Phi(h)^*:=\Phi(\overline{h}).
		\end{equation}
		The choice for real test functions is here made in accordance with Ref.\ \cite{kayTheoremsUniquenessThermal1991}.}
	\begin{itemize}
		\item \textit{(linearity)} for $a\in\mathbb{R}$ and $f,g\in\mathcal{D}(M)$ we have 
		\begin{equation}
			\Phi(af+g)=a\Phi(f)+\Phi(g),
		\end{equation}
		\item \textit{(hermiticity)} for $f\in\mathcal{D}(M)$ we have
		\begin{equation}
			\Phi(f)^*=\Phi(f),
		\end{equation}
		\item \textit{(dynamics)} for $f\in\mathcal{D}(M)$ and $P:\mathcal{E}(M)\rightarrow\mathcal{E}(M)$ the differential operator defined in eq.\ \eqref{eq:kg} we have
		\begin{equation}
			\Phi(Pf)=0,
		\end{equation}
		\item \textit{(CCR)} for $f,g\in\mathcal{D}(M)$ we have
		\begin{equation}
			[\Phi(f),\Phi(g)]=\textup{i}E(f,g).
		\end{equation}
	\end{itemize}
	Here $E:\mathcal{D}(M)^2\rightarrow\mathbb{R}$ is the commutator function defined by
	\begin{equation}
	\label{eq:comm}
	E(f,g)=\int_M \text{dvol}\, gG_c(f),
	\end{equation}
	with $G_c=G_--G_+$ the causal propagator and $G_\pm$ the unique advanced and retarded propagators given by $PG_\pm(f)=f$ and $\text{supp}(G_\pm(f))\subset J^\pm(\text{supp}(f))$ (see Ref.\ \cite{barWaveEquationsLorentzian2007}). \\
	
	\noindent A state $\omega:\mathcal{A}(M)\rightarrow\mathbb{C}$, which for hermitian operators can be interpreted as mapping the operator onto its expectation value, is a normalized (i.e. $\omega(1)=1$) positive semi-definite (i.e. $\omega(aa^*)\in\mathbb{R}_{\geq0}$) complex-linear map. In this work we restrict ourselves to quasi-free (or Gaussian) states, defined uniquely by their two-point functions 
		\begin{equation}
			\Lambda(f,g):=\omega(\Phi(f)\Phi(g)),
		\end{equation}
	typically taken to be (complex valued) bi-distributions on $\mathcal{D}(M)$, where all higher order $n$-point functions are calculated via Isserlis theorem (see Ref.\ \cite{khavkineAlgebraicQFTCurved2015}). As reviewed in the same reference, such a two point function must have the properties
	\begin{itemize}
		\item $\Lambda\circ(P\otimes1)=\Lambda\circ(1\otimes P)=0,$
		\item $Re\circ\Lambda$ is a positive semi-definite symmetric real bilinear map on $\mathcal{D}(M)^2$,
		\item $Im\circ\Lambda=\frac{1}{2}E$,
		\item $\vert E(f,g)\vert^2\leq 4\Lambda(f,f)\Lambda(g,g).$
	\end{itemize}
	In fact any bi-distribution satisfying the properties above defines a quasi-free state.
	
	Quasi-free states are said to satisfy the (local) Hadamard condition if on any Cauchy normal neighbourhood $U\subset M$ the two-point function has an asymptotic expansion of a `fundamental solution' to $P_x\Lambda(x,y)=0$
	\begin{equation}
	\label{eq:had}
	\Lambda\sim\frac{1}{8\pi^2}\left[\frac{u}{\sigma_+}+\sum_{n=0}^\infty v^{(n)}\sigma^n\ln\left(a^{-2}\sigma_+\right)+w^{(n)}\sigma^n\right].
	\end{equation}
	Here $\sigma$ is the Synge world function, $a$ an arbitrary length scale and $u,v^{(n)},w^{(n)}$ are smooth functions satisfying certain transport equations, which can be derived from $P\Lambda(x,y)=0$ (see Ref.\ \cite{dewittRadiationDampingGravitational1960} or Sec.\ \ref{sec:char_had}). The (local) bi-distribution $\frac{1}{\sigma_+}$ (and similarly $\ln(\sigma_+)$) is defined on $\mathcal{D}(U)$ as
	\begin{equation}
		\left(\frac{1}{\sigma_+}\right)(f,g)=\lim_{\varepsilon\downarrow0}\int_U\text{dvol}_x\int_U\text{dvol}_y\frac{f(x)g(y)}{\sigma(x,y)+\textup{i}a\varepsilon(\mathcal{T}(x)-\mathcal{T}(y))+a^2\varepsilon^2},
	\end{equation}
	where $\mathcal{T}$ is an arbitrary time-function on $M$. In the following sections we shall often write this as
	\begin{equation}
		\sigma_+(f,g)=\sigma(f,g)+\textup{i}0^+(\mathcal{T}(x)-\mathcal{T}(y))+\left(0^+\right)^2.
	\end{equation}
	A more complete discussion on this definition, including many subtleties that are not addressed here, can be found in Ref.\ \cite{kayTheoremsUniquenessThermal1991}. Here it should be noted that some technical details in the extension of the local definition above to global Hadamard states, defined on a region larger than a Cauchy normal neighbourhood, have been addressed only recently in Ref.\ \cite{morettiGlobalHadamardParametrix2021}.
	
	As first shown in Ref.\ \cite{radzikowskiMicrolocalApproachHadamard1996}, the Hadamard condition can also be formulated in the language of microlocal analysis (see Appendix \ref{app:micro}). A two-point function is Hadamard if and only if it satisfies the microlocal spectrum condition, i.e. for its wave front set we have
	\begin{equation}
		WF'(\Lambda)=\left\{((x_1,k_1),(x_2,k_2))\in (T^*(M)\setminus\mathbf{0})^2:(x_1,k_1)\sim(x_2,k_2),k_1\in \left(V_{x_1}^+\right)^d\right\},
	\end{equation}
	where $\left(V_{x_1}^+\right)^d$ is the dual of the closed future light cone.
	This alternative characterization has proven very useful for studying general properties of Hadamard states, and can be nicely generalized beyond the scalar field in 3+1 dimensions, as seen for instance in Ref.\ \cite{sahlmannMicrolocalSpectrumCondition2001}. While we will not make direct use of tools from microlocal analysis, in later sections we definitely rely on results on Hadamard states derived from this reformulation. Still, we deem a detailed discussion of the wave front set mentioned above beyond the scope of this text. In essence, the wave front set captures aspects of the singular ultraviolet behaviour of a (bi-)distribution, generalizing the singular support of a distribution. It is important to keep in mind that any arbitrary bi-distribution satisfying the microlocal spectrum condition will generally not have the asymptotic expansion give in eq.\ \eqref{eq:had}. As theorem 5.1 in Ref.\ \cite{radzikowskiMicrolocalApproachHadamard1996} indicates, to really establish the two-way connection between this expansion and the microlocal spectrum condition, one needs the equation of motion $\Lambda\circ(P\otimes1)=\Lambda\circ(1\otimes P)=0$ and the commutation relation $Im(\Lambda)=\frac{1}{2}E$, or rather, that these identities hold up to smooth integral kernel. Hence a bi-distribution satisfying the microlocal spectrum condition, and 
	\begin{equation}
		WF'(\Lambda\circ(P\otimes1))=WF'(\Lambda\circ(1\otimes P))=WF'\left(Im(\Lambda)-\frac{1}{2}E\right)=\emptyset,
	\end{equation}
	can be referred to as a Hadamard parametrix. Notably, the fact that a two-point function is positive semi-definite is not required in the theorem cited above. In fact, theorem 6.3 of the same reference tells us that positivity (again modulo a smooth function) is automatically satisfied for a Hadamard parametrix.
	
	The most well-known example of a Hadamard state is the Minkowski vacuum state for free theories on flat space-time, but in fact any ground- or thermal (i.e. KMS) state on a static space-time satisfies the microlocal specrum condition (see Ref.\ \cite{sahlmannPassivityMicrolocalSpectrum2000} and \cite{sandersThermalEquilibriumStates2013}). A state satisfying the Hadamard condition in a neighbourhood of a Cauchy surface will satisfy the Hadamard condition on the whole space-time, hence the existence of Hadamard states on any globally hyperbolic space-time can be asserted from existence on static space-times using a deformation argument (see Ref.\ \cite{fullingSingularityStructureTwopoint1981}). One can also give existence results via a direct construction as performed in Ref.\ \cite{lewandowskiHadamardStatesBosonic2020}.
	
	\subsubsection{The boundary theory}	
	\label{sec:qft_bound_setup}
	For each $T\in\mathbb{R}$ we can consider the algebra $\mathcal{A}_T:=\mathcal{A}(M;M_T)$.\footnote{For any open set $U\subset M$, the localized algebra $\mathcal{A}(M;U)\subset\mathcal{A}(M)$ is defined as the smallest unital *-algebra such that for any $f\in\mathcal{D}(U)$ (understood as a subset of $\mathcal{D}(M)$) we have $\Phi(f)\in\mathcal{A}(M;U)$.} Clearly, these algebras satisfy $\mathcal{A}_T\subset \mathcal{A}_{T'}$ for any $T'\leq T$. Following Ref.\ \cite{gerardConstructionHadamardStates2016} we can also associate an algebra $\mathcal{B}_T$ with the characteristic surface $C_T$ such that $\mathcal{A}_T\cong\mathcal{B}_T$, which we view as a bulk to boundary correspondence, in the way described below. 
	
	Recalling the definition of $R$ in eq.\ \eqref{eq:R} and that of the causal propagator $G_c$ in section \ref{sec:qft_bulk_setup}, we define the map $\rho_T:\mathcal{D}(M_T)\rightarrow\tilde{\mathcal{S}}_{s.c.}(C_T)\subset\mathcal{E}(C_T)$ via
	\begin{equation}
		\rho_T(f)=RG_c(f)\restriction_{C_T},
	\end{equation}
	where $\tilde{\mathcal{S}}_{s.c.}(C_T)$ is defined such that this map is surjective. As proven in Ref.\ \cite{baerInitialValueProblems2015}, 
	\begin{equation}
		\rho_T(f)=\rho_T(g)\text{ implies }G_c(f)\restriction_{M_T}=G_c(g)\restriction_{M_T},
	\end{equation} and hence, by theorem 3.4.7 in Ref.\ \cite{barWaveEquationsLorentzian2007}, $f-g\in P\mathcal{D}(M_T)$. This means that $\rho_T:\mathcal{D}(M_T)/P\mathcal{D}(M_T)\rightarrow\tilde{\mathcal{S}}_{s.c.}(C_T)$ is a linear isomorphism.
	
	We can define a symplectic structure $\sigma_{C_T}$ on $\tilde{\mathcal{S}}_{s.c.}(C_T)$ via
	\begin{equation}
		\sigma_{C_T}(F,G)=2\int_0^\infty \diff r\int_{S^2}\diff\Omega \, F\partial_r G.
	\end{equation}
	This is chosen such that the map $\rho_T$ becomes a symplectomorphism, as using the form of the Klein-Gordon equation in eq.\ \eqref{eq:kgchar}, we can see that for $f,g\in\mathcal{D}(M_T)$
	\begin{align}
	E(f,g)=	&\int_{M_T}\text{dvol}\,gG_c(f)\nonumber\\
	=	&2\int_T^\infty \diff t\int_0^\infty \diff r\int_{S^2}\diff\Omega \,RG_c(f)\left[-\partial_t\partial_r-K_t\right]RG_-(g)\nonumber\\
	=	&2\int_0^\infty \diff r\int_{S^2}\diff \Omega \,\left[RG_c(f)\partial_rRG_-(g)\right]\restriction_{C_T}+\int_{M_T}\text{dvol}\,G_-(g)PG_c(f)\nonumber\\
	=	&2\int_0^\infty \diff r\int_{S^2}\diff \Omega \,\rho_T(f)\partial_r\rho_T(g)\nonumber\\
	=	&\sigma_{C_T}(\rho_T(f),\rho_T(g)),
	\end{align}
	where we have used that $K_T$ is symmetric on $\tilde{\mathcal{S}}_{s.c.}(C_T)$ when given the $L^2(C_T,\diff r\diff\Omega)$ inner product, that near $r=0$ we have $RG_c(g)=\mathcal{O}(r)$ and that $RG_c(g)$ is spatially compact, i.e. there is an $\tilde{r}>0$ such that $\text{supp}(RG_c(g)\restriction_{C_T})\subset\{r<\tilde{r}\}$.\\
	
	\noindent Completely analogously to as in section \ref{sec:qft_bulk_setup}, we can now construct an algebra $\mathcal{B}_T$ generated by elements $\Psi(F)$ for $F\in\tilde{\mathcal{S}}_{s.c.}(C_T)$ satisfying
	\begin{itemize}
		\item $\Psi(aF+G)=a\Psi(F)+\Psi(G)$,
		\item $\Psi(F)^*=\Psi(F)$,
		\item $[\Psi(F),\Psi(G)]=\textup{i}\sigma_{C_T}(F,G)$.
	\end{itemize}
	The symplectomorphism $\rho_T$ induces a natural *-isomorphism $\iota_T:\mathcal{A}_T\rightarrow\mathcal{B}_T$ via 
	\begin{equation}
		\iota_T(\Phi(f))=\Psi(\rho_T(f)).
	\end{equation} 
	Note that while the dynamics of the theory is implemented in (the net structure of) $\mathcal{A}_T$, we cannot say the same for $\mathcal{B}_T$. The latter theory can only be seen as dynamical by virtue of its bulk to boundary correspondence. However, as we will see in section \ref{sec:char_dyn}, we can also understand this dynamics in terms of the family of algebras $\{\mathcal{B}_T:T\in\mathbb{R}\}$ and their relations.\\
	
	\noindent Clearly, just as on $\mathcal{A}_T$, we can define quasi-free states on $\mathcal{B}_T$ using the 2-point function $\lambda_T:\tilde{\mathcal{S}}_{s.c.}(C_T)^2\rightarrow\mathbb{C}$ satisfying
	\begin{itemize}
		\item $Re\circ\lambda_T$ is a positive definite symmetric real bilinear map,
		\item $Im\circ\lambda_T=\frac{1}{2}\sigma_{C_T}$,
		\item $\vert\sigma_{C_T}(F,G)\vert^2\leq 4\lambda_T(F,F)\lambda_T(G,G)$.		
	\end{itemize}
	A two-point function $\lambda_T$ can be used to define a two-point function $\Lambda_T$ on $\mathcal{D}(M_T)$ via 
	\begin{equation}
	\label{eq:boundtobulk_2pt}
	\Lambda_T(f,g)=\lambda_T(\rho_T(f),\rho_T(g)).
	\end{equation}
	Reversely, given a two-point function $\Lambda$  on $\mathcal{D}(M)$, we can give a formal expression for $\lambda_T$ such that $\Lambda_T=\Lambda\restriction_{\mathcal{D}(M_T)^2}$ via
	\begin{equation}
	\label{eq:bulktobound_2pt}
	\lambda_T(r,\Omega;r',\Omega')=4\partial_r\partial_{r'}R(T,r)R(T,r')\Lambda(T,r,\Omega;T,r',\Omega').
	\end{equation}
	Here $\Lambda(x;x')$ is the (formal) integration kernel of $\Lambda$ (with respect to the volume measure induced by the metric) and $\lambda_T(r,\Omega;r',\Omega')$ the formal integration kernel of $\lambda_T$ (with respect to the (hypersurface) area element $\diff r\diff \Omega$). This relation can be derived from noting that for $f,g\in\mathcal{D}(M_T)$ 
	\begin{equation}
		\Lambda(f,g)=\int_{M_T}\text{dvol}_xf(x)\Lambda(x,g)=\sigma_{C_T}(R\Lambda(.,g)\restriction_{C_T},\rho_T(f)).
	\end{equation}
	Of course this only holds if $\Lambda(.,g)$ is sufficiently regular. As also noted in Ref.\ \cite{kayTheoremsUniquenessThermal1991}, the formal relation \eqref{eq:bulktobound_2pt} cannot be made precise for arbitrary distributions. Luckily, we can make sense of this expression for Hadamard states. That is to say, as we will see in section \ref{sec:char_had}, to do this one needs an appropriate choice of time-function in the definition of $\frac{1}{\sigma_+}$.\\
	
	\noindent As was shown in Ref.\ \cite{gerardConstructionHadamardStates2016}, a two-point function $\lambda_T$ induces a Hadamard two-point function $\Lambda_T$ on $\mathcal{D}(M_T)$ if it satisfies the \textit{characteristic micro-local spectrum condition} (c$\mu$SC): 
	\begin{itemize}
		\item ${}_{C_t}WF'(\lambda_t)=WF'(\lambda_t)_{C_t}=\emptyset,$
		\item $WF'(\lambda_t)\cap\{((r,\Omega),(\sigma,\xi);(r',\Omega'),(\sigma',\xi'))\in T^*(C_t\times C_t): \sigma>0\text{ or }\sigma'>0\}=\emptyset,$
	\end{itemize}
	where we identify $T^*_{(r,\Omega)}C_t\cong \mathbb{R}\times T^*_\Omega S^2$. Here we again point to appendix \ref{app:micro} for the relevant definitions and notation regarding the wave front set. Notably, these conditions do not fully fix the singular behaviour of $\lambda_T$, i.e. for two boundary two-point functions, $\lambda_T$ and $\lambda_T'$, satisfying the c$\mu$SC, their difference $\lambda_T-\lambda_T'$ need not be smooth. This is contrast with what we know for Hadamard two-point functions on the bulk. If we consider $\Lambda_T$ and $\Lambda_T'$, their difference $\Lambda_T-\Lambda_T'$ has a smooth integral kernel on $M_T^2$. However, this kernel may generally not be extendible as a smooth function to $M^2$ (and hence may not define a smooth function on $C_T^2$ via equation \eqref{eq:bulktobound_2pt}). Here lies the reason that the c$\mu$SC on $\lambda_T$ doesn't fix the singular behaviour, as this condition does not guarantee that $\Lambda_T$ can be extended as a Hadamard two-point function to all of $M$, or at the very least to a neighbourhood of $C_T$. Physically, this has as a consequence that for these states the expectation values of observables such as the stress-energy tensor may blow up as one approaches $C_T$ from above. Contrastingly, suppose that $\lambda_T$ and $\lambda_T'$ define two-point functions $\Lambda_T$ and $\Lambda_T'$ that \textit{can} be extended as Hadamard bi-distributions to all of $\mathcal{D}(M)$, we have in particular that 
	\begin{equation}
		(\Lambda_T-\Lambda_T')\restriction_{C_T^2}(r,\Omega;r',\Omega'):=\lim_{t\downarrow T}(\Lambda_T-\Lambda_T')(t,r,\Omega;t,r',\Omega')
	\end{equation} exists and is a smooth function. This in turn implies that
	\begin{equation}
		(\lambda_T-\lambda_T')(r,\Omega;r',\Omega')=4\partial_r\partial_{r'}RR'(\Lambda_T-\Lambda_T')\restriction_{C_T^2}(r,\Omega;r',\Omega')
	\end{equation}
	is smooth. Hence $\lambda_T$ and $\lambda_T'$ have the same singular behaviour. It is this class of boundary two-point functions that we shall refer to as \textit{characteristic Hadamard two-point functions}.\\
	
	\noindent The characteristic Hadamard two-point functions shall be the main object of study in section \ref{sec:char_had}. Arguably the most immediate question is what the singular behaviour, given for instance in terms of an asymptotic expansion analogous to eq.\ \eqref{eq:had}, is for these states.\footnote{In principle one could consider feeding eq.\ \eqref{eq:had} directly into eq.\ \eqref{eq:bulktobound_2pt}. However, this would typically not give a globally defined parametrix, nor would the restriction of $\sigma$ to $C_T^2$, at least on a neigbourhood where it is defined, be a convenient object to work with.} This question is relevant if one wants to consider the semi-classical Einstein equations in terms of a characteristic initial value problem. Here one considers some initial state on a characteristic surface such that at least the stress-energy tensor is well-defined at this surface. One method to determine this singular behaviour would be via an explicit calculation of $\lambda_T$ from some global Hadamard two-point function $\Lambda$ using \eqref{eq:bulktobound_2pt}. This is done for the vacuum state of a massless field on Minkowski space-time in Ref.\ \cite{kayTheoremsUniquenessThermal1991}. However for general spherically symmetric space-times this can be rather tedious, as it for instance requires us to know the behaviour of $\sigma$, the Synge world function, and of $u$ in a neighbourhood of the null geodesics that rule $C_T$. In this paper we take an alternative approach that, arguably, allows us to determine the necessary singular behaviour of $\lambda_T$ from first principles. We will see in section \ref{sec:char_had} that the main benefit of this approach is that it automatically yields us a \textit{characteristic Hadamard parametrix} $h_T$, i.e. a bi-distribution defined globally on $C_T$ such that for each characteristic Hadamard function $\lambda_T$ the difference $\lambda_T-h_T$ is smooth. As we will discuss in section \ref{sec:SET}, this yields a point splitting procedure to renormalize operators such as the stress-energy tensor with some particularly nice behaviour (compared to directly point splitting with regards to a Hadamard parametrix that is locally constructed). In particular, this procedure generalizes the point-splitting procedure on conformally flat space-times discussed in Ref.\ \cite{pinamontiConformalGenerallyCovariant2009}.
	
	\section{The characteristic Hadamard parametrix}
	\label{sec:char_had}
	The asymptotic expansion of Hadamard states in eq.\ \eqref{eq:had} allows one to locally define (order $N$) Hadamard parametrices
	\begin{equation}
		H_N=\frac{1}{8\pi^2}\left[\frac{u}{\sigma_+}+\sum_{n=0}^Nv^{(n)}\sigma^n\ln(a^{-2}\sigma_+)\right],
	\end{equation}
	such that $\Lambda-H_N$ has a $2N+1$ times continuously differentiable distribution kernel, which modulo a $C^{2N+1}$ contribution is uniquely determined by requiring that $H_N\circ(P\otimes1)$ is $2N-1$ times continuously differentiable and $u(x,x)=1$.\footnote{Strictly speaking, these are not Hadamard parametrices in the sense of how we described them in section \ref{sec:qft_bulk_setup}, as the $P$ acting on $H_N$ generally does not yield a smooth function, however for many purposes, especially when defining non-linear observables of finite order, such an `approximate' notion of a Hadamard parametrix is sufficient.} In particular, following Ref.\ \cite{dewittRadiationDampingGravitational1960}, or for a more recent and timely account, Ref.\ \cite{frobTraceAnomalyChiral2019},
	one can use the defining relations of the Synge world function 
	\begin{equation}
		\nabla_\mu\sigma\nabla^\mu\sigma=2\sigma,\; \sigma(x,x)=0,
	\end{equation}
	where the covariant derivatives are understood to act on the first coordinate, to derive that $H_N$ is a parametrix if $u$ and $v^{(n)}$ satisfy the following \textit{transport equations}
	\begin{equation}
		2\nabla^\mu\sigma\nabla_\mu\ln(u)+\Box\sigma-4=0,
	\end{equation}
	\begin{equation}
		v^{(0)}+\nabla^\mu\sigma\left(\nabla_\mu v^{(0)}-v^{(0)}\nabla_\mu\ln(u)\right)+\frac{1}{2}Pu=0,
	\end{equation}
	\begin{equation}
		(n+2)v^{(n+1)}+\nabla^\mu\sigma\left(\nabla_\mu v^{(n+1)}-v^{(n+1)}\nabla_\mu\ln(u)\right)+\frac{1}{2(n+1)}Pv^{(n)}=0.
	\end{equation}
	Given $u(x,x)=1$ this system of equations is uniquely solved on Cauchy normal neighbourhoods, in particular we get $u=\sqrt{\Delta}$ with $\Delta$ the van Vleck-Morette determinant. Furthermore, it has been shown in Ref.\ \cite{morettiProofSymmetryOffdiagonal2000} that the functions $u$ and $v^{(n)}$, the latter of which are referred to as the Seeley-DeWitt (or sometimes Hadamard) coefficients, are symmetric.\footnote{\label{foot:Hambig}In actual fact, there is some ambiguity in the choice of transport equations, or more generally in the definition of the Seeley-deWitt coefficients. For instance, the function $u$ can in principle be modified freely by any $\mathcal{O}(\sigma)$ contribution. However, for any Hadamard parametrix $H_N$ with the desired properties, these ambiguities only lead to $2N+1$-times continuously differentiable differences in the definition and are hence often irrelevant. The coefficients defined by the transport equations above are seen as the canonical choice, though of course even in this case, the bi-distributions $H_N$ still contain a further ambiguity in the choice of reference scale $a$.} A further important feature of these coefficients, and hence of $H_N$, is that they are locally covariant. That is, for fixed mass $m$ and curvature coupling $\xi$, $u(x,x')$ and $v^{(n)}(x,x')$ only depend on the geometry of a geodesically convex neighbourhood of $x$ and $x'$. This is why the order $N$ Hadamard parametrices can be used to give locally covariant definitions of non-linear observables for this quantum field theory, such as the stress-energy tensor, via a point-splitting procedure (see Ref.\ \cite{hollandsLocalWickPolynomials2001}).\\
	
	\noindent We wish to do an analogous computation to construct a characteristic Hadamard parametrix in terms of functions that one can naturally define on a characteristic hypersurface.\footnote{For two points on a characteristic surface that are not placed on the same bicharacteristic, i.e. the null curves generated by the vectorfield $e_1$, the functions $\sigma$, $u$ and $v_n$ depend highly non-trivially on the surrounding geometry of the surface. We aim to find an expansion describing the same singular behaviour that is more adapted to our characteristic approach.} Recall that in order for the $\mu$SC to imply the Hadamard condition for a bi-distribution, we need to impose a dynamical requirement, i.e. that it satisfies the equation of motion up to a smooth source, and that it is consistent with the (singular part of the) commutation relation. Comparing this to the c$\mu$SC, the commutation relation can be imposed at the boundary, and the dynamical requirement inside the bulk is imposed via the bulk to boundary correspondence. However, we do not have a dynamical requirement at the boundary itself, the bulk two-point function need not satisfy the equation of motion at a point on the characteristic boundary (or, as two-point functions are distributional, rather on `test functions' that are non-zero at some boundary submanifold). Hence, in order to derive the singular structure of characteristic Hadamard two-point functions, we shall impose a condition on these boundary two-point functions that is related to the equation of motion being satisfied at this boundary. We implement this dynamical requirement on boundary two-point functions via the evolution of these two-point functions over the family of characteristic surfaces $\{C_t:t\in\mathbb{R}\}$, as outlined in the following section.
	
	\subsection{The dynamics of boundary two-point functions}
	\label{sec:char_dyn}
	Suppose that $\Lambda$ is a Hadamard two-point function on $M$. Let $\lambda_t$ be the family of boundary two-point functions on $C_t$ such that 
	\begin{equation}
		\Lambda\restriction_{\mathcal{D}(M_t)^2}=\Lambda_t=\lambda_t\circ(\rho_t\otimes\rho_t).
	\end{equation} For $f,g\in\mathcal{D}(M_t)$ we can always find a $t'> t$ such that $f,g\in\mathcal{D}(M_{t'})$. This means that $\Lambda_t(f,g)=\Lambda_{t'}(f,g)$. In particular
	\begin{equation}
		\partial_t\Lambda_t(f,g)=0.
	\end{equation}
	Using the bulk to boundary correspondence, we find that this implies
	\begin{equation}
		0=\partial_t\Lambda_t(f,g)=\dot\lambda_t(\rho_t(f),\rho_t(g))+\lambda_t(\dot\rho_t(f),\rho_t(g)))+\lambda_t(\rho_t(f),\dot\rho_t(g)))=0.
	\end{equation}
	Due to the fact that $G_c(f)$ has compact spatial support, we can integrate the classical equation of motion \eqref{eq:kgchar} to write
	\begin{equation}
		\dot\rho_t(f)(r,\Omega)=\int_r^\infty \diff s \,(K_t\rho_t(f))(s,\Omega).
	\end{equation}
	Formally, we can use this to write down a dynamical equation for the integration kernel $\lambda_t(r,\Omega;r',\Omega')$, namely
	\begin{align}
	\label{eq:char_ker_dyn}
	\dot\lambda_t(r,\Omega;r',\Omega')+K_t\int_0^r\diff s\,\lambda_t(s,\Omega;r',\Omega')+K'_t\int_0^{r'}\diff s\,\lambda_t(r,\Omega;s,\Omega')=0,
	\end{align}
	where $K_t$ should be interpreted as acting on the unprimed coordinates (and be dependent on $r$) and ${K'_t}$ on the primed coordinates, and depending on $r'$.\footnote{See appendix \ref{app:dervs} for a derivation of this equation.}\\
	
	\noindent On our class of spherically symmetric space-times with quasi-conformal coordinates as given in Sec.\ \ref{eq:metric}, we now wish to find a family of bi-distributions $h_t$ on $\mathcal{S}_{s.c.}(C_t)$ that satisfy the c$\mu$SC, the commutation relation and such that
	\begin{equation}
	\label{eq:char_par_dyn}
	\dot h_t+h_t\circ\left(O_t\otimes 1 \right)+h_t\circ\left(1\otimes O_t\right)=-4S_t,
	\end{equation}
	for some family of functions $S_t\in \mathcal{E}(M^2)\restriction_{C_t^2}$ smooth in $t$ and where the map $O_t:\mathcal{S}_{s.c.}(C_t)\rightarrow C^\infty(C_t)$ is given by
	\begin{equation}
	O_t(f)(r,\Omega)=\int_r^\infty \diff s \,(K_tf)(s,\Omega).
	\end{equation}
		
	\noindent Given some fixed $T$, we now consider a bulk two-point function $\Lambda_T=\lambda_T\circ(\rho_T\otimes\rho_T)$ such that its corresponding boundary two-point function satisfies 
	\begin{equation}
		\lambda_T=h_T+4\partial_r\partial_r'RR'w_T,
	\end{equation} with 
	\begin{equation}
		w_T\in\mathcal{E}(M^2)\restriction_{C_T^2}.
	\end{equation}
	This bulk two-point function $\Lambda_T$ defines a family of boundary two point functions $\{\lambda_t\}_{t\geq T}$, via the state induced on $\mathcal{B}_t\cong\mathcal{A}_t\subset\mathcal{A}_T$. We now claim that
	\begin{equation}
		\lambda_t-h_t\in\partial_r\partial_{r'}RR'\mathcal{E}(M^2)\restriction_{C_t^2},
	\end{equation}
	and hence we can find a $w_t\in\mathcal{E}(M^2)\restriction_{C_t^2}$ with 
	\begin{equation}
		w_t(r,\Omega;r',\Omega')=\frac{1}{4RR'}\int_0^r\diff s\int_0^{r'}\diff{s'}\,(\lambda_t-h_t)(s,\Omega;s',\Omega).
	\end{equation}
	In particular, such a $w_t$ should satisfy
	\begin{equation}
	\label{eq:wtkg}
	\left(\partial_t\partial_r\partial_{r'}+K_t\partial_{r'}+K'_t\partial_r\right)RR'w_t=S_t.
	\end{equation}
	Indeed, given a $w_t$ satisfying equation \eqref{eq:wtkg} for $t>T$, with $w_T$ determined from $\Lambda_T$, then we can see that the family $\lambda_t=h_t+4\partial_r\partial_{r'}RR'w_t$ satisfies $\partial_t\Lambda_t(f,g)=0$ for any $f,g\in\mathcal{D}(M_t)$. Now we can easily conclude that
	\begin{equation}
		\Lambda_t(f,g)=\Lambda_T(f,g).
	\end{equation}
	The question that remains, is whether eq.\ \eqref{eq:wtkg} indeed has smooth solutions. To show this, we write 
	\begin{equation}
		w_t(r,\Omega;r',\Omega')=w_{t,T}+\int_T^t\diff s\,F_{t,s}(r,\Omega;r',\Omega'),
	\end{equation}
	where for $t>T$ 
	\begin{equation}
		\left(\partial_t\partial_r\partial_{r'}+K_t\partial_{r'}+K'_t\partial_r\right)RR'w_{t,T}=0,
	\end{equation}
	and for $t>s\geq T$
	\begin{equation}
		\left(\partial_t\partial_r\partial_{r'}+K_t\partial_{r'}+K'_t\partial_r\right)RR'F_{t,s}=0,
	\end{equation}
	such that  
	\begin{equation}
		\lim_{t\downarrow T} w_{t,T}=w_T,
	\end{equation}		
	and 
	\begin{equation}
		\lim_{t\downarrow s}F_{t,s}=\frac{1}{RR'}\int_0^r\diff r\int_0^{r'}\diff r'S_s.
	\end{equation}
	For this latter boundary condition, it is important to note that $\frac{1}{RR'}\int_0^r\diff r\int_0^{r'}\diff r'S_s$, as well as all of its derivatives in $r$ and $r'$, can be continuously extended to $r=0$ and $r'=0$.
	
	For $t>s\geq T$, both $w_{t,T}$ and $F_{t,s}$ exist and are smooth, as they can be given as restrictions to $C_t^2$ of the (smooth) integral kernel of
	\begin{equation}
		4\int_0^\infty \diff r\int_{S^2}\diff \Omega\int_0^\infty \diff r'\int_{S^2}\diff\Omega'\,\rho_T(f)(r,\Omega)\rho_T(g)(r',\Omega')\partial_{r}\partial_{r'}RR'w_{T}(r,\Omega;r',\Omega'),
	\end{equation} and 
	\begin{equation}
		4\int_0^\infty \diff r\int_{S^2}\diff \Omega\int_0^\infty \diff r'\int_{S^2}\diff\Omega'\,\rho_s(f)(r,\Omega)\rho_s(g)(r',\Omega')S_s(r,\Omega;r',\Omega'),
	\end{equation}
	respectively.\footnote{That these integral kernels are smooth, can be seen from simple analogues to the proof of theorem 5.3 in Ref.\ \cite{gerardConstructionHadamardStates2016}, where in this case we have for the initial data that $WF'(\partial_r\partial_{r'}RR'w_T)=WF'(S_s)=\emptyset$.}
	To show that $w_t$ exists and is smooth, we first note that $F_{t,s}$ is continuous as a function of $s$. After all, we have that
	$\partial_{s}F_{t,s}$ is the restriction to $C_t^2$ of the integration kernel of
	\begin{align}
	-&\int_0^\infty \diff r\int_{S^2}\diff\Omega\int_0^\infty \diff r'\int_{S^2}\diff\Omega'\,\rho_s(f)(r,\Omega)\rho_s(g)(r',\Omega')\times\nonumber\\
	&\left[\dot S_s(r,\Omega;r',\Omega')+K\int_0^r\diff r\,S_s(r,\Omega;r',\Omega')+K'\int_0^{r'}\diff r'\,S_s(r,\Omega;r,\Omega')\right].
	\end{align}
	Therefore, as $\partial_{s}F_{t,s}$ exists as a smooth function of $(r,\Omega;r',\Omega')$ for all $s<t$, we know that $F_{t,s}$, as well as all its derivatives in $r$, $r'$, $\Omega$ and $\Omega'$, are continuous as a function of $s$. Furthermore, as both $F_{t,t}$ and $F_{t,T}$ exist as smooth functions on $C_t^2$, then for all fixed $(r,\Omega;r',\Omega')$, we know that $F_{t,s}$ is bounded on $s\in[T,t]$ and hence (Riemann) integrable. Hence we know that $w_t$ exists and that furthermore, by the Leibniz integral rule, we know that $w_t$ is smooth.\\
	
	\noindent To construct a characteristic Hadamard parametrix satisfying the dynamical condition \eqref{eq:char_par_dyn} in the same vain as how one could construct $H_N$, we wish to make an Ansatz about the form of $h_t$. To understand which Ansatz to make, it is helpful to consider a simple class of space-times and quantum field theories on which the full singular structure of characteristic Hadamard states is known explicitly.
	
	\subsubsection{A relevant example: the conformal vacuum}
	On a conformally flat spherically symmetric space-time (i.e. $\alpha=0$), we consider a massless conformally coupled scalar field ($m=0$, $\xi=\frac{1}{6}$). For such a theory we can define the conformal vacuum state (see for instance Ref.\ \cite{pinamontiConformalGenerallyCovariant2009}) via
	\begin{multline}
	\label{eq:conf_vac_bulk}
	\Lambda(t,r,\Omega;t',r',\Omega')=\\\frac{1}{4\pi^2}\left[\frac{\exp(-\beta(t,r)-\beta(t',r'))}{-(t-t')^2-2(t-t'-\textup{i}0^+)(r-r'-\textup{i}0^+)+2rr'(1-\cos(\theta))}\right],
	\end{multline}
	where $\theta$ is the relative angle between $\Omega$ and $\Omega'$.\footnote{Note that the time-function that is chosen here to regularize the integral kernel of this state is given by $\mathcal{T}=r+t$, which is indeed time-like as \begin{equation}g(\nabla\mathcal{T},\nabla\mathcal{T})=-\exp(-2\beta).\end{equation} This temporal function is chosen such that in the calculation of $\lambda_t$, one need not worry about time-splitting, i.e. doing the calculation on $C_t\times C_{t+0^+}$, in order for \eqref{eq:bulktobound_2pt} to yield a well-defined distribution. When one simply chooses $\mathcal{T}=t$, the calculation of $\lambda_t$ would yield an expression for the integration kernel where integrals diverge in the limit $\varepsilon\downarrow0$. This technicality is discussed in Ref.\ \cite{kayTheoremsUniquenessThermal1991}.} Notably on Minkowski space-time this just matches the Minkowski vacuum state
	\begin{equation}
		\Lambda(t,r,\Omega;t',r',\Omega')=\frac{1}{8\pi^2\sigma_{+}}.
	\end{equation}
	Doing a calculation completely analogous to the derivation of eq.\ (B.53) in Ref.\ \cite{kayTheoremsUniquenessThermal1991}, we see that the conformal vacuum boundary two-point functions are given by
	\begin{equation}
	\label{eq:conf_vac}
	\lambda_t(r,\Omega;r',\Omega')=\frac{-1}{\pi}\frac{\delta_{S^2}(\Omega,\Omega')}{(r-r'-\textup{i}0^+)^2}.
	\end{equation}
	Fur such a boundary two-point function, one can indeed show that eq.\ \eqref{eq:char_ker_dyn} holds.
	
	\subsection{An Ansatz for the singular behaviour of characteristic Hadamard states}
	If we consider an arbitrary scalar field on a non-conformally flat space-time, the two-point functions of eq.\ \eqref{eq:conf_vac} still satisfies the c$\mu$SC and is has an imaginary part that yields the correct commutation relations. Hence, it defines a Hadamard two-point functions $\Lambda_t$ on $M_t$. In fact, for states of this form this was already proven in Ref.\ \cite{hollandsstefanAspectsQuantumField2000}. If we calculate $\partial_t\Lambda_t$ for this family of states (see Appendix \ref{app:dervs}), we find that
	\begin{equation}
		\label{eq:leading_source}
		\partial_t\Lambda_t(f,g)=\frac{1}{\pi}\int \diff r\diff r'\diff\Omega\, \rho_t(f)(r,\Omega)\rho_t(g)(r',\Omega)\kappa_t^{(0)}(r,r'),
	\end{equation}
	with
	\begin{align}
	\label{eq:kappa0}
	2\kappa^{(0)}_t(r,r'):=&\frac{\exp(2\beta_t(r'))\tilde{V}(r')-\exp(2\beta_t(r))\tilde{V}(r)}{r'-r}\nonumber\\
	&+\frac{r\exp(2\beta_t(r))\tilde{V}(r)+r'\exp(2\beta_t(r'))\tilde{V}(r')}{rr'}\nonumber\\
	&-\frac{2\left(\exp(2\alpha(r'))-\exp(2\alpha(r))\right)-\left(\partial_r\exp(2\alpha(r))+\partial_{r'}\exp(2\alpha(r'))\right)(r'-r)}{(r'-r)^3}.
	\end{align}
	In particular, we see that the boundary integral kernel of $\partial_t\Lambda_t$ is singular, as it contains a $\delta_{S^2}$-like behaviour.\footnote{At first glance it may seem surprising that there are no terms containing $\Delta_\Omega\delta_{S^2}$ appearing in this distribution. After all, the operator $K$ contains a term involving the operator $\Delta_\Omega$. However, as we see in Appendix \ref{app:dervs}, this term indeed vanishes upon close inspection.	The fact that this term does not contribute for a relatively simple two-point function, is a feature of our chosen of coordinates. For more general forms of the metric, one sees that the leading order singular term of the characteristic Hadamard expansion has to be modified into a more complicated form to get the same result.}
	Aside from this singular behaviour in the angular coordinates, we can see that for smooth space-times, $\kappa_t^{(0)}$ is in fact smooth at $r=r'>0$ (or rather has a unique smooth extension to this submanifold). That being said, in general this $\kappa^{(0)}$ still diverges near $r=0$ or $r'=0$. We wish to use eq.\ \eqref{eq:conf_vac} as the leading order singularity in our Ansatz, so we will need to find additional lower order singular terms to ensure that $\partial_t\Lambda_t$ has a smooth boundary integral kernel. Analogous to the bulk Hadamard expansion, we make the following Ansatz,
	\begin{equation}
	\label{eq:char_had_ans}
	\lambda_t(r,\Omega;r',\Omega')=\frac{-1}{\pi}\frac{\delta(\Omega,\Omega')}{(r-r'-\textup{i}0^+)^2}+\frac{1}{2\pi^2}\partial_r\partial_{r'}R(t,r)R(t,r')\tilde{\lambda}_t(r,\Omega;r',\Omega'),
	\end{equation}
	with 
	\begin{equation}
		\tilde{\lambda}_t=w_t+v_t\ln(\frac{RR'(1-\cos\theta)}{a^2}).
	\end{equation}
	Here we assume that $w_t,v_t\in\mathcal{E}(M^2)\restriction_{C_T^2}$.
	Note that the logarithmic divergences precisely coincide with pairs of points on $C_t$ that are located on a bicharacteristic. This is why it is reasonable to expect that these logarithmic divergences match those of $\Lambda$ when restricted to $C_t^2$, i.e. that $v_t$ is related to the Hadamard coefficients $\{v^{(n)}:n\in\mathbb{N}\}$. We will verify this explicitly in section \ref{sec:renorm}. If we consider purely our dynamical conditions on characteristic Hadamard states, the appearance of these logarithmic divergences is also natural, as here we see that 
	\begin{equation}
		\Delta_\Omega\ln(1-\cos\theta)=4\pi\delta_{S^2}(\Omega,\Omega')-1.
	\end{equation}
	It is this relation that allows us to cancel all singularities appearing in $\partial_t\Lambda_t$ when $v_t$ is chosen such that it satisfies certain transport equations.
	\subsection{The transport equations for the characteristic Hadamard coefficients}
	Similarly to the bulk Hadamard states, $v_t$ should be state-independent and geometrical. That means in particular that it should respect the symmetries of the space-time, which enables us to formally expand $v_t$ as
	\begin{equation}
		v_t(r,\Omega;r',\Omega')=\sum_{n=0}^\infty v_t^{(n)}(r,r')(1-\cos\theta)^n.
	\end{equation}
	In principle, we don't know a priori if this expansion is absolutely convergent, but at least one can interpret this as an asymptotic expansion. Notably, unlike the asymptotic expansion in eq.\ \eqref{eq:had}, this expansion is unique.\footnote{Recall our comment on the ambiguity of $v^{(n)}$ in footnote \ref{foot:Hambig}.} We shall refer to $v_t^{(n)}$ as the \textit{characteristic Hadamard coefficients}. \\
	
	\noindent This expansion allows us to rewrite the condition $\partial_t\Lambda_t=0$ as
	\begin{align}
	0=&\partial_r\partial_{r'}\partial_tRR'\tilde{\lambda}_t-\frac{1}{2}\Delta_\Omega\left[\frac{R}{r^2}\partial_{r'}R'\tilde{\lambda}_t+\frac{R'}{r'^2}\partial_{r}R\tilde{\lambda}_t\right]\nonumber\\
	&+\frac{1}{2}\left[\left(\exp(2\beta)\tilde{V}-\partial_r\exp(2\alpha)\partial_r\right)R\partial_{r'}R'\tilde{\lambda}_t\right.\nonumber\\
	&+\left.\left(\exp(2\beta')\tilde{V}'-\partial_{r'}\exp(2\alpha')\partial_{r'}\right)R'\partial_{r}R\tilde{\lambda}_t\right]\nonumber\\
	&+2\pi\kappa^{(0)}_t\delta_{S^2}\nonumber\\
	=&2\pi\left(\kappa^{(0)}-2\frac{R'}{r'^2}\partial_{r}Rv_t^{(0)}\right)\delta_{S^2}+\nonumber\\
	&\sum_{n=0}^\infty(n+1)^2(1-\cos\theta)^n\ln(a^{-2}RR'(1-\cos\theta))\Big[\kappa_t^{(n+1)}-2\frac{R'}{r'^2}\partial_{r}Rv_t^{(n+1)}\Big]\nonumber\\
	&+(1-\cos\theta)^n\Big[\frac{1}{RR'}\partial_tRR'\partial_r\partial_{r'}RR'v_t^{(n)}+\partial_t((\partial_rR)(\partial_{r'}R')v_t^{(n)}-RR'\partial_r\partial_{r'}v_t^{(n)})\nonumber\\
	&+\frac{R'}{r'^2}\partial_{r}R\left((2n+1)v_t^{(n)}-4(n+1)v_t^{(n+1)}\right)\nonumber\\
	&+\frac{R'}{r'^2}(\partial_{r}R)\left(n(n+1)v_t^{(n)}-2(n+1)^2v_t^{(n+1)}\right)\nonumber\\
	&+\left(\exp(2\beta)\tilde{V}-\partial_r\exp(2\alpha)\partial_r\right)R(\partial_{r'}R')v_t^{(n)}\nonumber\\
	&-\partial_r\exp(2\alpha)(\partial_rR)\partial_{r'}R'v_t^{(n)}-\frac{(\partial_rR)}{R}\exp(2\alpha)\partial_rR\partial_{r'}R'v_t^{(n)}\Big]\nonumber\\
	&+(\partial_t\partial_r\partial_{r'}+K_t\partial_{r'}+K'_t\partial_r)RR'\tilde{w}_t+(r\leftrightarrow r'),
	\end{align}
	with
	\begin{align}
	\label{eq:higher_kappa}
	\kappa_t^{(n+1)}=&\frac{1}{2(n+1)^2}\Big[\partial_t\partial_{r}\partial_{r'}RR'v_t^{(n)}+\left(\exp(2\beta)\tilde{V}-\partial_r\exp(2\alpha)\partial_r\right)R\partial_{r'}R'v_t^{(n)}\nonumber\\
	&+(n+1)n\frac{R'}{r'^2}\partial_{r}Rv_t^{(n)}\Big]+(r\leftrightarrow r').
	\end{align}
	We can read off the transport equation
	\begin{equation}
	\label{eq:lambda_recursion}
	\left[\frac{R}{r^2}\partial_{r'}R'+\frac{R'}{r'^2}\partial_{r}R\right]v_t^{(n)}=\kappa_t^{(n)},
	\end{equation}
	which has the unique solution (demanding $v_t^{(n)}$ to be finite at $r=0$ and $r'=0$)
	\begin{equation}
	v_t^{(n)}(r,r')=\frac{(rr')^2}{RR'}\int_{0}^\infty \diff s\, \frac{\kappa^{(n)}\left(\frac{r}{1+rs},\frac{r'}{1+r's}\right)}{(1+rs)^2(1+r's)^2}.
	\end{equation}
	We arrive at the conclusion that for characteristic Hadamard states, the asymptotic expansion of $v_t$ is uniquely determined by the geometry near the lightcone $C_t$. The remaining smooth part $w_t$ is state-dependent and satisfies the dynamical equation
	\begin{equation}
	\label{eq:smooth_dyn}
	(\partial_t\partial_r\partial_{r'}+K_t\partial_{r'}+K'_t\partial_r)RR'w_t=S_t
	\end{equation}
	with
	\begin{align}
	S_t=-\frac{1}{2}\sum_{n=0}^\infty(1-\cos\theta)^n\Big[&\frac{1}{RR'}\partial_tRR'\partial_{r}\partial_{r'}RR'v_t^{(n)}\nonumber\\&+\partial_t\left((\partial_rR)(\partial_{r'}R')v_t^{(n)}-RR'\partial_r\partial_{r'}v_t^{(n)}\right)\nonumber\\
	&+\frac{R'}{r'^2}\partial_{r}R\left((2n+1)v_t^{(n)}-4(n+1)v_t^{(n+1)}\right)\nonumber\\
	&+\frac{R'}{r'^2}(\partial_{r}R)\left(n(n+1)v_t^{(n)}-2(n+1)^2v_t^{(n+1)}\right)\nonumber\\
	&+(\partial_{r'}R')\left(\exp(2\beta)\tilde{V}-\partial_r\exp(2\alpha)\partial_r\right)Rv_t^{(n)}\nonumber\\
	&-\frac{1}{R}(\partial_rR)\exp(2\alpha)\partial_r\partial_{r'}RR'v_t^{(n)})\nonumber\\
	&-\partial_r\exp(2\alpha)(\partial_rR)\partial_{r'}R'v_t^{(n)})\nonumber\\
	&+(r\leftrightarrow r')\Big].\label{eq:source}
	\end{align}
	
	\noindent Note that $S_t$ is independent of the reference scale $a$ introduced in the logarithmic part of the two-point function. Hence one sees that for each Hadamard distribution where the smooth part $w_t$ satisfies Eq.\ \eqref{eq:smooth_dyn}, one directly finds a whole family of Hadamard bi-solutions parameterized by $a$. However, it should be noted that these distributions in general do not define a two-point function, as for arbitrary reference scale $a$ this map may not be positive semi-definite.
	
	For general space-times and couplings the source term above is somewhat unwieldy. It is likely not possible to write it in a closed form, or even find a comprehensive formula for each term in the expansion. Luckily, calculating the expansion of the source-term above to some finite order is still useful when wanting to calculate certain coincidence limits of $w_t$, as discussed in Sec.\ \ref{sec:hawking_rad}. 
	Nevertheless, in some simple cases we can find formulas for the terms in the expansion to arbitrary order, for which we shall give a non-trivial example of below.
	\subsection{Example: the massive field on Minkowski space-time}
	\label{sec:massive}
	On Minkowski space-time, i.e. $\alpha=\beta=0$, we consider a scalar field with mass $m$. In particular, this means that $\tilde{V}=m^2$. For quasi-free states satisfying the Ansatz \eqref{eq:char_had_ans} we find that
	\begin{equation}
		\kappa_t^{(0)}=\frac{m^2}{2}\left(\frac{1}{r}+\frac{1}{r'}\right).
	\end{equation}
	This implies that $v_t^{(0)}=\frac{m^2}{2}=v^{(0)}$, where $v^{(0)}$ is the Seeley-DeWitt coefficient given by the transport equations from the beginning of this section. This is a good first indication that our findings are consistent with the known expansion of the Hadamard singularity. More generally, using also that our space-time is time-translation invariant, we see that our recurrence relation given by \eqref{eq:higher_kappa} and \eqref{eq:lambda_recursion} simplify to
	\begin{align}
	\kappa_t^{(n+1)}=\frac{1}{2(n+1)^2}\Bigg[&\left((m^2-\partial_r^2)\partial_{r'}+(m^2-\partial_{r'}^2)\partial_{r}\right)rr'v_t^{(n)}\nonumber\\&+n(n+1)\left(\frac{1}{r^2}\partial_{r'}+\frac{1}{r'^2}\partial_r\right)rr'v_t^{(n)}\Bigg],
	\end{align}
	\begin{equation}
		\left[\frac{1}{r^2}\partial_{r'}+\frac{1}{r'^2}\partial_r\right]rr'v_t^{(n)}=\kappa_t^{(n)}.
	\end{equation}
	Making the further Ansatz
	\begin{equation}
		v_t^{(n)}=\frac{m^2}{2}a_n(m^2rr')^n,
	\end{equation}
	we calculate 
	\begin{equation}
		\kappa^{(n+1)}=\frac{m^2}{2}\frac{a_n}{2(n+1)}(m^2rr')^{n+1}\left(\frac{1}{r}+\frac{1}{r'}\right),
	\end{equation}
	and see that the recurrence relation reduces to
	\begin{equation}
		a_n=\frac{a_{n-1}}{2(n+1)n},
	\end{equation}
	hence
	\begin{equation}
		a_n=\frac{1}{2^nn!(n+1)!}.
	\end{equation}
	
	In fact, we can use the dynamical equation \eqref{eq:smooth_dyn} to (almost fully) specify the smooth part of the vacuum state. We note that for the source term $S_t$ we have
	\begin{align}
	S_t=\left(\frac{1}{r}+\frac{1}{r'}\right)\frac{m^2}{2}\sum_{n=0}^\infty(m^2rr')^{n+1}(1-\cos\theta)^n\frac{2n+3}{2^{n+1}n!(n+2)!}.
	\end{align}
	The vacuum state ought to be invariant under continuous Poincar\'e transformations. This means in particular that in the bulk two-point function the only dependence on space-time points enters via the Synge world function $\sigma$ and (time)-orientation, where the latter is of particular relevance in the commutator. The Hadamard coefficients $v_t^{(n)}$ found above define a (bulk) Hadamard parametrix that indeed respects these symmetries. This means that the smooth part of the two-point function ought to do the same. Therefore the function $w_t$ is a function of $\sigma\restriction_{C_t^2}=rr'(1-\cos\theta)$, hence we make the Ansatz
	\begin{equation}
		w_t=\frac{m^2}{2}\sum_{n=0}^\infty b_n(m^2rr'(1-\cos\theta))^n,
	\end{equation}
	we find that \eqref{eq:smooth_dyn} yields
	\begin{equation}
		\left(\frac{1}{r}+\frac{1}{r'}\right)\frac{m^2}{2}\sum_{n=0}^\infty(m^2rr')^{n+1}(1-\cos\theta)^n\frac{n+1}{2}(b_n-2(n+1)(n+2)b_{n+1})=S_t,
	\end{equation}
	from which one can deduce the relation
	\begin{equation}
		b_{n+1}=\frac{b_n}{2(n+1)(n+2)}-a_{n+1}\frac{2n+3}{(n+1)(n+2)}.
	\end{equation}
	Note that this system of equations needs an initial condition on $b_0$ to be uniquely solvable. We can then express all coefficients $b_n$ in terms of $b_0$,
	\begin{equation}
		b_n=\frac{1}{2^nn!(n+1)!}\left[b_0-\sum_{j=1}^n\frac{2j+1}{j(j+1)}\right]=\frac{1}{2^nn!(n+1)!}\left(b_0-2H[n]+\frac{n}{n+1}\right),
	\end{equation}
	where $H[n]=\sum_{j=1}^n\frac{1}{j}$ are the harmonic numbers. Using the convention $H[0]=0$, we can now write down the full 2-point function of the massive vacuum state
	\begin{align}
	\Lambda_m=\frac{1}{8\pi^2}\Bigg[\frac{1}{\sigma_+}+\frac{m^2}{2}\sum_{n=0}^\infty\frac{(m^2\sigma)^n}{2^nn!(n+1)!}\left(\ln(\frac{\sigma_+}{a^2})-2H[n]+\frac{n}{n+1}\right)\Bigg],
	\end{align}
	where we have absorbed the unknown $b_0$ in to the reference scale $a$. Recalling the remark made below Eq.\ \eqref{eq:source}, the reference scale $a$ still needs to be fixed to a particular value such that the bi-distribution above defines a positive semi-definite two-point function. One can show that the expression above can be written in terms of modified Bessel functions. Let $F_c=K_1+cI_1\in\mathcal{E}(\mathbb{R}_{>0})$, with $I_1$ and $K_1$ modified Bessel functions of the first and second kind respectively and $c\in\mathbb{R}$ an a priori free parameter, such that $F_c$ solves the modified Bessel equation
	\begin{equation}
	x^2\partial_x^2F_c(x)+x\partial_xF_c(x)-(x^2+1)F_c(x)=0.
	\end{equation}
	We can now rewrite the expression for $\Lambda_m$ above as
	\begin{align}
		\Lambda_m=\frac{m}{4\pi^2\sqrt{2\sigma^+}}F_c(m\sqrt{2\sigma_+}),
	\end{align}
	where using the known expansions for the modified Bessel functions for $x\downarrow 0$ (see Ref.\ \cite[p.\ 375]{abramowitz+stegun})
	\begin{equation}
	I_1(x)=\frac{1}{2}x+\mathcal{O}(x^2),
	\end{equation}
	and
	\begin{equation}
		K_1(x)=\frac{1}{x}+I_1(x)\ln(x)+\frac{2\gamma-\ln(4)-1}{4}x+\mathcal{O}(x^2),
	\end{equation}
	where $\gamma$ is the Euler-Masceroni constant, we can relate
	\begin{equation}
		c=\frac{1}{2}\ln(\frac{e^{2\gamma-1}m^2}{2a^2}).
	\end{equation}
	Of course the two-point function of the proper Minkowski vacuum state is well known, and exactly matches the case where $c=0$, or $a^2=\frac{e^{2\gamma-1}m^2}{2}$. To show that this is indeed the unique choice for $c$ such that $\Lambda_m$ is positive semi-definite, we rely on an argument involving the behaviour of the two-point function at large space-like separation. Write
	\begin{equation}
		\Lambda_m=\Lambda_{m,0}+cW,
	\end{equation}
	where
	\begin{equation}
		\Lambda_{m,0}=\frac{m}{4\pi^2\sqrt{2\sigma^+}}K_1(m\sqrt{2\sigma_+})
	\end{equation}
	the proper vacuum state two-point function and $W\in\mathcal{E}(M\times M)$ given by
	\begin{equation}
		W=\frac{m}{4\pi^2\sqrt{2\sigma}}I_1(m\sqrt{2\sigma}).
	\end{equation}
	Due to the exponential decay of $K_1(x)$ as $x\rightarrow\infty$, we know that $\Lambda_{m,0}$ satisfies the cluster decomposition property (see Ref.\ \cite{streater}). Let $f,g\in\mathcal{D}(\mathbb{R}^4)$ and $\tau_\mathbf{v}:\mathcal{D}(M)\rightarrow\mathcal{D}(M)$ a translation map for a space-like vector $\mathbf{v}$ given by 
	\begin{equation}
		\tau_{\mathbf{v}}f(\mathbf{x})=f(\mathbf{x}-\mathbf{v}),
	\end{equation}
	where we have identified points on Minkowski space with fourvectors. The cluster decomposition property for this Poincare invariant two-point functions tells us that
	\begin{equation}
		\lim_{s\rightarrow\infty}\Lambda_{m,0}(f+\tau_{s\mathbf{v}}g,f+\tau_{s\mathbf{v}}g)=\Lambda_{m,0}(f,f)+\Lambda_{m,0}(g,g).
	\end{equation}
	$I_1(x)$ on the other hand is known to exponentially diverge (to positive infinity) for $x\rightarrow\infty$. If we therefore choose $f\in\mathcal{D}(M)$ such that $f\geq 0$ and $f\not\in P\mathcal{D}(M)$, one easily sees that for each $B>0$ there is an $d>0$ such that for $s>d$
	\begin{equation}
		W(f,\tau_{s\mathbf{v}}f)>B.
	\end{equation}
	Therefore for $c\neq 0$, 
	\begin{equation}
		\Lambda_m(f-c\tau_{s\mathbf{v}}f,f-c\tau_{s\mathbf{v}}f)\rightarrow -\infty \text{ as }s\rightarrow \infty.
	\end{equation}
	we conclude that $c=0$ is the only possible choice to yield a positive definite two-point function. Hence, using our methods, we regain the expanded form of the vacuum two-point function for the massive scalar field on Minkowski space-time
	\begin{align}
	\Lambda_m=\frac{1}{8\pi^2}\Bigg[\frac{1}{\sigma_+}+\frac{m^2}{2}\sum_{n=0}^\infty\frac{(m^2\sigma)^n}{2^nn!(n+1)!}\left(\ln(\frac{\textup{e}^{2\gamma-1}m^2\sigma_+}{2})-2H[n]+\frac{n}{n+1}\right)\Bigg].
	\end{align}
		
	\noindent The state studied above is arguably very elementary and well understood, therefore this construction does not really reveal anything new about the massive vacuum state. Nevertheless, it acts as a good consistency check for the methods we have introduced. In the next section, we will show that the characteristic Hadamard parametrix
	\begin{equation}
		h_t=\frac{-1}{\pi}\frac{\delta_{S^2}(\Omega,\Omega')}{(r-r'-\textup{i}0^+)^2}+\frac{1}{2\pi^2}\partial_r\partial_{r'}RR'v_t\ln(a^{-2}RR'(1-\cos\theta)),
	\end{equation}
	although not a locally covariant object, can be a very useful tool when calculating the expectation values of non-linear observables, especially the stress-energy tensor.
	
	\section{Wick squares, the stress-energy tensor and all that}
	\label{sec:renorm}
	We shall now discuss how to use the characteristic parametrix defined in Section \ref{sec:char_had} to calculate expectation values of renormalized locally covariant non-linear observables such as Wick squares and the stress-energy tensor for a two-point function $\lambda_t$ on the boundary $C_t$. Naively, one might define non-linear observables by point-splitting $\lambda_t$ and subtracting the characteristic parametrix. However, such a procedure generally does not yield a locally covariant renormalization scheme. Instead we require that our observables are defined via a locally covariant renormalization scheme as described in \cite{hollandsLocalWickPolynomials2001}. In particular, the observables are constructed via a point-splitting procedure where a locally covariant Hadamard parametrix $H$ is subtracted. Here we also take note of the finite renormalization ambiguity due to the non-uniqueness of this prescription.\footnote{In the case of the stress-energy tensor some of this ambiguity is fixed by additional requirements, such as divergencelessness. See Ref.\ \cite{waldQuantumFieldTheory1994}.}
	We start by calculating a simple Wick square $\langle:\Phi^2:\rangle$, which for a bulk two-point function $\Lambda$ is defined as
	\begin{equation}
		\langle:\Phi^2:\rangle_\Lambda(x)=\lim_{x'\rightarrow x}(\Lambda-H)(x',x)+c_1m^2+c_2\mathfrak{R}(x),
	\end{equation}
	where $c_1,c_2\in\mathbb{R}$ denote the renormalization ambiguities (depending analytically on $\xi$).
	If we formally restrict a Hadamard two-point function $\Lambda$ to $C_t^2$, we now know that
	\begin{align}
	\label{eq:char2pt}
		\Lambda(t,r,\Omega;t,r',\Omega')=\frac{1}{8\pi^2}\Big[&\frac{\exp(-\beta(t,r)-\beta(t,r'))}{\textup{i}0^+(r-r'-\textup{i}0^+)+rr'(1-\cos(\theta))}\nonumber\\
		&+v_t(r,\Omega;r',\Omega')\ln\left(a^{-2}RR'(1-\cos\theta)\right)+w_t(r,\Omega;r',\Omega')\Big],
	\end{align}
	where we recognise the leading order term as the conformal vacuum 2-point function restricted to the cone.
	We can now use this to rewrite the Wick square as
	\begin{align}
	8\pi^2\langle:\Phi^2:\rangle_\Lambda(t,r,\Omega)=&w_t(r,\Omega;r,\Omega)\nonumber\\
	&+\lim_{\Omega'\rightarrow \Omega}\frac{1}{R^2(1-\cos(\theta))}-\frac{u(t,r,\Omega;t,r,\Omega')}{\sigma(t,r,\Omega;t,r,\Omega')}\nonumber\\
	&\begin{aligned}+\lim_{\Omega'\rightarrow \Omega}&v_t^{(0)}(r,r)\ln\left(a^{-2}R^2(1-\cos\theta)\right)\\&-v^{(0)}(t,r,\Omega;t,r,\Omega')\ln\left(a^{-2}\sigma(t,r,\Omega;t,r,\Omega')\right)\end{aligned}\nonumber\\
	&+8\pi^2\left(c_1m^2+c_2\mathfrak{R}(t,r,\Omega)\right).
	\end{align}
	To show that this is indeed finite, let us first observe that
	\begin{align}
	\sigma(t,r,\Omega;t,r,\Omega')=&[\partial_\theta^2\sigma](1-\cos\theta)+\frac{1}{6}\left([\partial_\theta^4\sigma]+[\partial_\theta^2\sigma]\right)(1-\cos\theta)^2+\mathcal{O}((1-\cos\theta)^3)\nonumber\\
	=&R^2(1-\cos\theta)\\&+\frac{1}{6}R^2(1-(\partial_rR)(\exp(2\alpha)(1+\partial_r\beta)-2r\partial_t\beta))(1-\cos\theta)^2+\mathcal{O}((1-\cos\theta)^3),
	\end{align}
	and
	\begin{align}
	u(t,r,\Omega;t,r,\Omega')=&[u]+[\partial_\theta^2u](1-\cos\theta)+\mathcal{O}((1-\cos\theta)^2)\nonumber\\
	=&1+\frac{1}{12}(d\Omega)^{\mu\nu}R_{\mu\nu}(1-\cos\theta)+\mathcal{O}((1-\cos\theta)^2),
	\end{align}
	where we have used the square bracket notation $[...]$ to denote the coincidence limit of a biscalars (see Ref.\ \cite{fullingAspectsQuantumField1989}).
	It follows that 
	\begin{equation}
		\lim_{\Omega'\rightarrow \Omega}\frac{1}{R^2(1-\cos(\theta))}-\frac{u(t,r,\Omega;t,r,\Omega')}{\sigma(t,r,\Omega;t,r,\Omega')}=\frac{[\partial_\theta^4\sigma]+[\partial_\theta^2\sigma]}{6R^4}-\frac{[\partial_\theta^2u]}{R^2},
	\end{equation}
	which is indeed finite.
	
	Turning our attention to the logarithmic divergences, we want to show that $[v^{(0)}_t]=[v^{(0)}]$, which we've already seen to hold for the massive field on Minkowski space-time. Indeed, if we expand the transport equation \eqref{eq:lambda_recursion} in powers of $(r'-r)$, equating lowest order terms gives
	\begin{equation}
	\partial_r R^2[v_t^{(0)}]=r^2[\kappa_t^{(0)}]=\frac{1}{2}\left(\partial_rR^2\tilde{V}+\frac{r^2}{6}\partial_r^3\exp(2\alpha)\right).
	\end{equation}
	Using relation \eqref{eq:vtilde} for $\tilde{V}$, this differential equation reduces to
	\begin{equation}
		\partial_r R^2[v_t^{(0)}]=\frac{1}{2}\partial_rR^2\left(m^2+\left(\xi-\frac{1}{6}\right)\mathfrak{R}\right)=\partial_r R^2[v^{(0)}],
	\end{equation}
	which indeed asserts that $[v^{(0)}_t]=[v^{(0)}]$.
	
	Putting this all together, we can finalize our computation of $\langle:\Phi^2:\rangle$, which gives
	\begin{equation}
	\langle:\Phi^2:\rangle_\Lambda\restriction_{C_t}=\frac{1}{8\pi^2}[w_t]+Q_{\Phi^2},
	\end{equation}
	with
	\begin{align}
	Q_{\Phi^2}=&\frac{(1-6[\partial_\theta^2u])[\partial_\theta^2\sigma]+[\partial_\theta^4\sigma]}{48\pi^2[\partial_\theta^2\sigma]^2}+\text{renormalization freedom}\nonumber\\=&c_1m^2+\left(c_2-\frac{1}{288\pi^2}\right)\mathfrak{R}-\frac{1}{48\pi^2}\mathbf{C}[e_0,e_1,e_0,e_1].
	\end{align}
	Here $\mathbf{C}$ is the Weyl tensor. It should be noted that while $Q_{\Phi^2}$ contains the locally covariant renormalization ambiguity, it is not as a whole locally covariant. In particular, the contribution 
	\begin{equation}
		\mathbf{C}[e_0,e_1,e_0,e_1]=-\frac{1}{3R^2}\left(1-\exp(2\alpha)+r\partial_r\exp(2\alpha)-\frac{1}{2}r^2\partial_r^2\exp(2\alpha)\right)
	\end{equation}		
	is a quantity associated with the local null tetrad depending on the chosen lightcone. Of course $[w_t]$ is also not a locally covariant quantity, as it is defined by subtracting the characteristic parametrix from $\lambda_t$. As we have defined $\langle:\Phi^2:\rangle$ such that it as a whole is locally covariant, we will refer to $Q_{\Phi^2}$ as the covariance correction term.
	
	As we will encounter the following terms in the stress-energy tensor, we will also calculate the covariance correction terms for 
	\begin{equation}
		\langle:\partial_r\Phi\partial_r\Phi:\rangle=\frac{1}{8\pi^2}[\partial_r\partial_{r'}w_t]+Q_{(\partial_r\Phi)^2},
	\end{equation}
	and 
	\begin{equation}
		\langle:\Vert\vec{\nabla}_{\Omega}\Phi\Vert^2_{S^2}:\rangle=\frac{1}{8\pi^2}[\nabla_\Omega\cdot\nabla_{\Omega'} w_t]+Q_{\Vert\vec{\nabla}_{\Omega}\Phi\Vert^2_{S^2}},
	\end{equation}
	which are calculated analogously to $Q_{\Phi^2}$.\footnote{To ease these calculations, we used the xAct packages for Mathematica (see Ref.\ \cite{j.m.martin-garciaet.al.XActEfficientTensor}). The notebooks involving these calculations can be made available upon request.} 
	We get
	\begin{align}
	Q_{(\partial_r\Phi)^2}=&-\frac{R_{rr}\left(m^2+\left(\xi-\frac{1}{6}\right)\mathfrak{R}\right)}{96\pi^2}-\frac{1}{2880\pi^2}\left(2\nabla_r\nabla_r\mathfrak{R}+\tensor{R}{_r^\mu} R_{\mu r}-\frac{8}{3}R_{rr}\mathfrak{R}\right)\nonumber\\
	&\begin{aligned}
	+\frac{1}{2880\pi^2}\Big(&-6\nabla^\alpha\nabla^\beta C_{\alpha r\beta r}-6\nabla_r\nabla_r\mathbf{C}[e_0,e_1,e_0,e_1]\\&+(7R_{rr}-9(\nabla_\alpha e_1^\alpha)^2(e_0)_r(e_0)_r)\mathbf{C}[e_0,e_1,e_0,e_1]\Big)
	\end{aligned}\nonumber\\
	&+\text{renormalisation freedom}
	\end{align}
	and
	\begin{align}
	Q_{\Vert\nabla_\Omega\Phi\Vert^2_{S^2}}=\left(d\Omega^2\right)^{\mu\nu}\Bigg[&\left(-\frac{R_{\mu\nu}\left(m^2+\left(\xi-\frac{1}{6}\right)\mathfrak{R}\right)}{96\pi^2}-\frac{1}{2880\pi^2}\left(2\nabla_\mu\nabla_\nu\mathfrak{R}+\tensor{R}{_\mu^\alpha} R_{\alpha \nu}-\frac{8}{3}R_{\mu\nu}\mathfrak{R}\right)\right)\nonumber\\
	&+g_{\mu\nu}\left(\frac{\mathfrak{R}\left(m^2+\left(\xi-\frac{1}{6}\right)\right)}{576\pi^2}-\frac{1}{17280\pi^2}\left(6R^{\alpha\beta\gamma\delta}R_{\alpha\beta\gamma\delta}-\mathfrak{R}^2-6R^{\alpha\beta}R_{\alpha\beta}\right)\right)\nonumber\\
	&+\frac{1}{2880\pi^2}\left(-6\nabla^\alpha\nabla^\beta C_{\alpha \mu\beta \nu}-6\nabla_\mu\nabla_\nu\mathbf{C}[e_0,e_1,e_0,e_1]+7R_{\mu\nu}\mathbf{C}[e_0,e_1,e_0,e_1]\right)\nonumber\\
	&\begin{aligned}
	+\frac{1}{960\pi^2}g_{\mu\nu}\Bigg(&\Box+20m^2+\left(10\xi-\frac{3}{2}\right)\mathfrak{R}-18\mathbf{C}[e_0,e_1,e_0,e_1]\\&+5\mathbf{R}[e_0,e_1,e_0,e_1]-6R^{-2}\Bigg)\mathbf{C}[e_0,e_1,e_0,e_1]\Bigg]
	\end{aligned}\nonumber\\
	&+\text{renormalisation freedom}
	\end{align}

	\subsection{The stress-energy tensor}
	\label{sec:SET}
	Classically, the stress energy tensor $\mathbf{T}$ of a linear scalar field (as defined for instance in Ref.\ \cite{hollandsConservationStressTensor2005}) is (in terms of its components) given by
	\begin{align}
	T_{\mu\nu}=&\nabla_\mu\phi\nabla_\nu\phi-\frac{1}{2}g_{\mu\nu}\nabla^\rho\phi\nabla_\rho\phi-\frac{1}{2}g_{\mu\nu}m^2\phi^2\nonumber\\&+\xi\left[R_{\mu\nu}-\frac{1}{2}g_{\mu\nu}\mathfrak{R}-\nabla_\mu\nabla_\nu+g_{\mu\nu}\Box\right]\phi^2.
	\end{align}
	This object is of main interest in the context of the (classical) Einstein equations
	\begin{equation}
	R_{\mu\nu}-\frac{1}{2}g_{\mu\nu}\mathfrak{R}=8\pi T_{\mu\nu}.
	\end{equation}
	Similarly, a quantum stress-energy tensor $:{\mathbf{T}}:$, which for general quantum field theories is related to their relative Cauchy evolution as given in Ref.\ \cite{brunettiGenerallyCovariantLocality2003}, can be used to write down the semi-classical Einstein equations
	\begin{equation}
	R_{\mu\nu}-\frac{1}{2}g_{\mu\nu}\mathfrak{R}=8\pi \langle : T_{\mu\nu}:\rangle,
	\end{equation}
	which dynamically couples a quantum field to a background geometry. The solution of such an equation is then given by a (globally hyperbolic) space-time $M$ and a state $\omega$ on $\mathcal{A}(M)$ such that $\langle : T_{\mu\nu}:\rangle$ is evaluated in $\omega$. The classical stress-energy tensor is divergenceless, and for the semi-classical Einstein equations to be consistent, this also needs to hold for the quantum stress-energy tensor,
	\begin{equation}
	\nabla^\mu \langle : T_{\mu\nu}:\rangle=0.
	\end{equation}
	Following the analysis in Ref.\ \cite{christodoulouProblemSelfgravitatingScalar1986}, we can see that on our class of spherically symmetric space-times there are only two functionally independent components of the stress-energy tensor, that is assuming that $\langle :{\mathbf{T}}:\rangle$ is smoothly defined on the entire space-time (which is the case if it is evaluated for a Hadamard state). Using our null tetrad introduced in Section \ref{sec:geom}, these independent components are $\langle :{\mathbf{T}}:\rangle[e_0,e_1]$ and $\langle :{\mathbf{T}}:\rangle[e_1,e_1]$. Classically these correspond to
	\begin{align}
	\mathbf{T}[e_0,e_1]=&\frac{1}{2r^2}\exp(-2\beta)\left(\frac{1}{2}\vec{\nabla}_\Omega^2\phi^2-\phi\vec{\nabla}_\Omega^2\phi\right)+\frac{1}{2}m^2\phi^2\nonumber\\&+e_0^\mu e_1^\nu\xi\left[R_{\mu\nu}-\frac{1}{2}g_{\mu\nu}\mathfrak{R}-\nabla_\mu\nabla_\nu+g_{\mu\nu}\Box\right]\phi^2,
	\end{align}
	and
	\begin{align}
	\mathbf{T}[e_1,e_1]=&\exp(-4\beta)\left(\frac{1}{2}\partial_r^2\phi^2-\phi\partial_r^2\phi\right)\nonumber\\
	&+e_1^\mu e_1^\nu\xi\left[R_{\mu\nu}-\frac{1}{2}g_{\mu\nu}\mathfrak{R}-\nabla_\mu\nabla_\nu+g_{\mu\nu}\Box\right]\phi^2.
	\end{align}
	In principle $:{\mathbf{T}}:$ is given in terms of Wick squares $:\Phi^2:$ and $:\Phi\nabla_\mu\nabla_\nu\Phi:$, and hence can be renormalized using a point-splitting procedure. However, not every renormalization scheme results in a stress-energy tensor that is divergenceless. Luckily, one can always transform a (locally covariant) renormalization scheme for the stress-energy tensor into a divergenceless definition by making use of the renormalization freedom of locally covariant Wick squares. Following Ref.\ \cite{waldQuantumFieldTheory1994}, or a more recent review in Ref.\ \cite{frobTraceAnomalyChiral2019}, if we denote the stress-energy tensor regularized by point-splitting and subtracting the Hadamard parametrix $H$ by ${\mathbf{T}}^\text{reg}$, the divergenceless quantum stress-energy tensor is given by
	\begin{equation}
	\langle: T_{\mu\nu}:\rangle=\langle  T_{\mu\nu}^\text{reg}\rangle+g_{\mu\nu}D+d_1K^{(1)}_{\mu\nu}+d_2K^{(2)}_{\mu\nu}+d_3m^2\left(R_{\mu\nu}-\frac{1}{2}g_{\mu\nu}\mathfrak{R}\right)+d_4m^4g_{\mu\nu},
	\end{equation}
	with $D$ the divergence correction
	\begin{align}
	D=\frac{1}{5760\pi^2}\Big(&2R^{\alpha\beta\gamma\delta}R_{\alpha\beta\gamma\delta}-2R^{\alpha\beta}R_{\alpha\beta}+5(1-6\xi)^2\mathfrak{R}^2\nonumber\\
	&+12(1-5\xi)\Box\mathfrak{R}-60m^2(1-6\xi)\mathfrak{R}+180m^4\Big),
	\end{align}
	$\mathbf{K}^{(i)}$ the conserved locally covariant geometric tensors of mass dimension 4
	\begin{align}
	K^{(1)}_{\mu\nu}=&-2\Box R_{\mu\nu}+\frac{2}{3}\nabla_\mu\nabla_\nu\mathfrak{R}+\frac{1}{3}g_{\mu\nu}\Box\mathfrak{R}\nonumber\\&-4R_{\alpha\mu\beta\nu}R^{\alpha\beta}+g_{\mu\nu}R_{\alpha\beta}R^{\alpha\beta}+\frac{4}{3}\mathfrak{R}R_{\mu\nu}-\frac{1}{3}g_{\mu\nu}\mathfrak{R}^2,
	\end{align}
	\begin{align}
	K^{(2)}_{\mu\nu}=2\nabla_\mu\nabla_\nu\mathfrak{R}-2g_{\mu\nu}\Box\mathfrak{R}-2\mathfrak{R}R_{\mu\nu}+\frac{1}{2}g_{\mu\nu}\mathfrak{R}^2,
	\end{align}
	and $d_i$ the remaining renormalization freedom.\footnote{When $\langle:\mathbf{T}:\rangle$ is used as a source in the semi-classical Einstein equations, one can absorb $d_3$ and $d_4$ into a redefinition of Newtons constant and an introduction of the Cosmological constant respectively. See Ref.\ \cite{waldQuantumFieldTheory1994}.}\\
	
	\noindent We want to express $\langle :{\mathbf{T}}:\rangle[e_0,e_1]$ and $\langle :{\mathbf{T}}:\rangle[e_1,e_1]$ on a characteristic surface $C_t$ in terms of the regularized boundary two-point function $w_t$, going through the same procedure as for the Wick squares, we find that
	\begin{align}
	\mathbf{T}[e_0,e_1]=&\frac{1}{8\pi^2}\Bigg(\frac{1}{2r^2}\exp(-2\beta)\left(\frac{1}{2}\vec{\nabla}_\Omega^2[w_t]-[\vec{\nabla}_{\Omega}^2w_t]\right)+\frac{1}{2}m^2[w_t]\nonumber\\&+e_0^\mu e_1^\nu\xi\left[R_{\mu\nu}-\frac{1}{2}g_{\mu\nu}\mathfrak{R}-\nabla_\mu\nabla_\nu+g_{\mu\nu}\Box\right][w_t]\Bigg)\nonumber\\
	&+\mathbf{Q}_{\mathbf{T}}[e_0,e_1],
	\end{align}
	and
	\begin{align}
	\mathbf{T}[e_1,e_1]=&\frac{1}{8\pi^2}\Bigg(\exp(-4\beta)\left(\frac{1}{2}\partial_r^2[w_t]-[\partial_{r}^2w_t]\right)\nonumber\\
	&+e_1^\mu e_1^\nu\xi\left[R_{\mu\nu}-\frac{1}{2}g_{\mu\nu}\mathfrak{R}-\nabla_\mu\nabla_\nu+g_{\mu\nu}\Box\right][w_t]\Bigg)\nonumber\\
	&+\mathbf{Q}_{\mathbf{T}}[e_1,e_1],
	\end{align}
	where
	\begin{align}
	\mathbf{Q}_{\mathbf{T}}[e_0,e_1]=&-\frac{1}{2r^2}\exp(-2\beta)Q_{\Phi\Delta_\Omega\Phi}+\frac{1}{2}m^2Q_{\Phi^2}\nonumber\\&+e_0^\mu e_1^\nu\xi\left[R_{\mu\nu}-\frac{1}{2}g_{\mu\nu}\mathfrak{R}-\nabla_\mu\nabla_\nu+g_{\mu\nu}\Box\right]Q_{\Phi^2}\nonumber\\&-D+r.f.,
	\end{align}
	and
	\begin{align}
	\mathbf{Q}_{\mathbf{T}}[e_1,e_1]=&\exp(-4\beta)\left(\frac{1}{2}\partial_r^2Q_{\Phi^2}-Q_{\Phi\partial_r^2\Phi}\right)\nonumber\\
	&+e_1^\mu e_1^\nu\xi\left[R_{\mu\nu}-\frac{1}{2}g_{\mu\nu}\mathfrak{R}-\nabla_\mu\nabla_\nu+g_{\mu\nu}\Box\right]Q_{\Phi^2}+r.f.,
	\end{align}
	with $r.f.$ the residual renormalization freedom. Generally the correction terms $\mathbf{Q}_{\mathbf{T}}[e_0,e_1]$ and $\mathbf{Q}_{\mathbf{T}}[e_1,e_1]$ contain second or higher order derivatives in $t$ of the functions $\alpha$ and $\beta$. However, in the case of conformal coupling ($\xi=\frac{1}{6}$), we can actually choose our renormalization freedom such that these terms do not appear. Setting 
	\begin{equation}
		d_2=\frac{1}{17280\pi^2},
	\end{equation}
	and using the tensor $\mathbf{H}$, defined in Ref.\ \cite{waldAxiomaticRenormalizationStress1978} as
	\begin{equation}
		H_{\mu\nu}=-\tensor{R}{_\mu^\alpha}R_{\alpha\nu}+\frac{2}{3}\mathfrak{R}R_{\mu\nu}+\frac{1}{2}R^{\alpha\beta}R_{\alpha\beta}g_{\mu\nu}-\frac{1}{4}\mathfrak{R}^2g_{\mu\nu},
	\end{equation}
	we find
	\begin{align}
	\mathbf{Q}_{\mathbf{T}}[e_0,e_1]=&\frac{1}{2880\pi^2}\mathbf{H}[e_0,e_1]-\frac{m^2}{96\pi^2}\mathbf{G}[e_0,e_1]+\frac{m^4}{32\pi^2}\mathbf{g}[e_0,e_1]\nonumber\\&\begin{aligned}+\frac{\left(d\Omega^2\right)^{\mu\nu}}{1920\pi^2R^2}\Bigg[&\frac{1}{3}\left(-6\nabla^\alpha\nabla^\beta C_{\alpha \mu\beta \nu}+4\nabla_\mu\nabla_\nu\mathbf{C}[e_0,e_1,e_0,e_1]-3R_{\mu\nu}\mathbf{C}[e_0,e_1,e_0,e_1]\right)\\
	&\begin{aligned}+g_{\mu\nu}\left(\frac{8}{3}\Box+\frac{1}{6}\mathfrak{R}-18\mathbf{C}[e_0,e_1,e_0,e_1]+5\mathbf{R}[e_0,e_1,e_0,e_1]-6R^{-2}\right)\times\\
	\mathbf{C}[e_0,e_1,e_0,e_1]\Bigg]\end{aligned}\end{aligned}\nonumber\\&+r.f.,
	\end{align}
	and
	\begin{align}
	\mathbf{Q}_{\mathbf{T}}[e_1,e_1]=&\frac{1}{2880\pi^2}\mathbf{H}[e_1,e_1]-\frac{m^2}{96\pi^2}\mathbf{R}[e_1,e_1]\nonumber\\
	&\begin{aligned}
	+\frac{1}{2880\pi^2}e_1^\mu e_1^\nu\Big(&-6\nabla^\alpha\nabla^\beta C_{\alpha \mu\beta \nu}+4\nabla_\mu\nabla_\nu\mathbf{C}[e_0,e_1,e_0,e_1]\\&-(3R_{\mu\nu}+9(\nabla_\alpha e_1^\alpha)^2(e_0)_\mu(e_0)_\nu)\mathbf{C}[e_0,e_1,e_0,e_1]\Big)
	\end{aligned}\nonumber\\
	&+r.f..
	\end{align}
	As mentioned above, if the state on which we are evaluating the stress-energy tensor is rotationally invariant, we can calculate all other components of the stress-tensor from these two. Given conformal coupling, we can use the fact that for this renormalization choice, the trace of the stress-energy tensor (involving the trace anomaly as found in Ref.\ \cite{waldTraceAnomalyConformally1978}) can be calculated to be 
	\begin{equation}
	\langle: T:\rangle=g^{\mu\nu}\langle: T_{\mu\nu}:\rangle=-m^2\langle:\Phi^2:\rangle+\frac{m^4}{32\pi^2}+\frac{1}{2880\pi^2}\left(R^{\alpha\beta\gamma\delta}R_{\alpha\beta\gamma\delta}-R^{\alpha\beta}R_{\alpha\beta}\right).
	\end{equation}
	This gives us
	\begin{align}
	\langle:{\mathbf{T}}[e_2,e_2]:\rangle=\langle:{\mathbf{T}}[e_3,e_3]:\rangle=\frac{1}{2}\left(g^{\mu\nu}\langle: T_{\mu\nu}:\rangle+2\langle: {\mathbf{T}}[e_0,e_1]:\rangle\right).
	\end{align}
	To find $\langle:{\mathbf{T}}[e_0,e_0]:\rangle$ we use the divergenceless of $\langle:{\mathbf{T}}:\rangle$. In particular we use
	\begin{equation}
		\langle: T_{\mu\nu}:\rangle\nabla^\mu e_0^\nu=\eta^{ab}\nabla_\mu e_a^\mu \langle:{\mathbf{T}}[e_b,e_0]:\rangle,
	\end{equation}
	where we note that 
	\begin{equation}
		\tensor{\nabla}{^{(\mu}} \tensor{e}{_0^{\nu)}}=\frac{1}{2r}\left(\exp(2\alpha)\left(r\partial_r\alpha-1\right)\tensor{e}{_0^{(\mu}}\tensor{e}{_1^{\nu)}}-\left(\exp(2\alpha)\left(r\partial_r\beta+1\right)-2r\partial_t\beta\right)g^{\mu\nu}\right).
	\end{equation}
	This allows us to write down
	\begin{align}
	\partial_rR^2\langle:{\mathbf{T}}[e_0,e_0]:\rangle=&-R^2\exp(2\beta)\langle: T_{\mu\nu}:\rangle\tensor{\nabla}{^{(\mu}} \tensor{e}{_0^{\nu)}}\nonumber\\&-\partial_tR^2\exp(2\beta)\langle:{\mathbf{T}}[e_0,e_1]:\rangle+\frac{1}{2}\partial_rR^2\exp(2(\alpha+\beta))\langle:{\mathbf{T}}[e_0,e_1]:\rangle,
	\end{align}
	So indeed given that we know $\langle:\mathbf{T}[e_0,e_1]:\rangle$ and $\langle:T:\rangle$ on some time interval, we can determine $\langle:\mathbf{T}[e_0,e_0]:\rangle$ (given that our geometry and state is sufficiently regular at $r=0$).
	Note that in the massless conformally flat limit we recover exactly the results from Ref.\ \cite{waldAxiomaticRenormalizationStress1978}.
	\section{An application: the response of non-linear observables to gravitational collapse}
	\label{sec:hawking_rad}
	As was first shown in Ref.\ \cite{hawkingParticleCreationBlack1975}, a black hole background gives rise to a particle production process for the quantum fields propagating on it. That is to say, for an asymptotically flat astrophysical black hole space-time, given that near asymptotic past null infinity the field is in the vacuum state (see Ref.\ \cite{dappiaggiHadamardStatesLightlike2017} where this notion is made precise), an observer far away from the black hole will at late times observe thermal radiation coming from the direction of the black hole.\footnote{As is well known, for an eventually stationary black hole, this late time radiation should be at the Hawking temperature 
	\begin{equation}
		T_H=\frac{\kappa}{2\pi},
	\end{equation}
	with $\kappa$ the surface gravity at the event horizon. In the context of black hole thermodynamics this temperature is associated to the black hole itself (see Ref.\ \cite{waldQuantumFieldTheory1994}), and in fact one can assign a similar temperature to any system with a bifurcate Killing horizon (see Ref.\ \cite{kayTheoremsUniquenessThermal1991}). In the case of the particle creation process mentioned above, the thermality of the radiation at late times can be related to the scaling limit of Hadamard states near the event-horizon of the black hole via a gravitation redshift effect (see Ref.\ \cite{fredenhagenDerivationHawkingRadiation1990} where this is made precise). In fact, even for black holes that are not asymptotically static, one can generalise these ideas with respect to the scaling limit near the apparent horizon of a black hole (see Ref.\ \cite{kurpiczTemperatureEntropyareaRelation2021}). As a consequence of these scaling limit arguments, one sees that the thermal behaviour is universal for any initial state, i.e. not just the asymptotic vacuum state, as long as it is Hadamard.} Such outgoing radiation carries away energy to infinity and, under some assumptions (see e.g. Ref.\ \cite{medaEvaporationFourdimensionalDynamical2021}), one can show that if the black hole background satisfies the semi-classical Einstein equations, this means that the black hole effectively loses mass, i.e. it has a shrinking apparent horizon. This effect is known as \textit{black hole evaporation}. If this effect continues until the horizon hits a black hole singularity, this gives rise to a number of conceptual issues that are often referred to as the \textit{information loss paradox}.\footnote{Information loss in this case means that for quantum fields on space-times with fully evaporating black holes, one expects that a final state (i.e. the state near future asymptotic infinity) has less information content (i.e. is less pure) than the initial state. This is at odds with the often held view that a fundamental theory of nature should be unitary and hence information loss ought not to occur in a theory underlying semi-classical gravity, i.e. quantum gravity (see Ref.\ \cite{marolfBlackHoleInformation2017}). This stance is not uncontroversial, as one can also take the point of view that information loss is just \textit{part of life} (see Ref.\ \cite{unruhInformationLoss2017}). These conceptual points aside, space-times of this sort also lead to more technical problems, as it is unclear if quantum field theories on space-times with such causal defects as the naked singularity produced at the end of evaporation even admit sensible states in the first place (see Ref.\ \cite{janssenQuantumFieldsSemiglobally2022}).} It should be noted that calculations on Hawking radiation observed at asymptotic infinity are of limited use when it comes to understanding the dynamics of semi-classical black hole formation and evaporation. One would like to know \textit{where and when} Hawking radiation (which includes both the late time thermal radiation as well as all other modes that one could observe at asymptotic future null infinity) is produced. Certainly its origin can be traced back to the gravitational collapse that ought to have formed a black hole, but in order to understand what the local backreaction of Hawking radiation is on the geometry as a whole, one needs to be able to locally calculate the stress-energy tensor of the quantum field, especially around the collapsing body. Analytic calculations are available in two-dimensional toy models (see Ref.\ \cite{juarez-aubryQuantumFieldsBlack2018}), but there is no guarantee that these results carry over to more realistic (four dimensional) models. We should also take note of numerical approaches to calculating the expectation values of locally covariant non-linear observables for quantum fields on black hole space-times, such as the methods described in Ref.\ \cite{leviVersatileMethodRenormalized2016} and in Ref.\ \cite{taylorModesumPrescriptionRenormalized2022}, as well as a lattice approach to semi-classical gravity recently proposed in Ref.\ \cite{bercziGravitationalCollapseQuantum2022}. However, in the former two approaches a full analysis of the stress-tensor of a quantum field on a dynamical black hole has, to the knowledge of the authors, not yet been carried out to satisfactory degree, while in the latter approach we feel some additional work is required to ensure that the stress tensor calculated using this method is (approximately) locally covariant in the relevant regimes.\\
	
	\noindent As an application of our characteristic approach to Hadamard states and the calculation of non-linear observables, we shall demonstrate how one can use these methods to calculate non-linear observables of Hawking radiation produced in a simple collapse model as an expansion in the black hole mass. In particular, we shall look at the response of the wick square $:\Phi^2:$ for a massless scalar field in the past asymptotic vacuum state, to the collapse of a thin shell of null dust. We shall also comment on how these methods can be generalized to calculate more involved non-linear observables on arbitrary (spherically symmetric) gravitational collapse space-times, though explicit calculations on these we shall leave for a future publication. Here it should be noted that, as the space-time associated with null shell collapse is not smooth and hence one has to be careful when applying results developed for smooth space-times. Nevertheless, we deem these calculations to be a promising starting point of an investigation to gain a better understanding of semi-classical black hole formation and evaporation.\\
	
	\noindent We consider a massless scalar field $\Phi:\mathcal{D}(M)\rightarrow\mathcal{A}(M)$, with $M$ a black hole space-time formed by collapsing null dust with a Schwarzschild radius $R_s$.\footnote{Note that since the Ricci scalar of a Vaidya space-time vanishes, the dynamics of the theory and the expectation values of wick squares evaluated for the past asymptotic vacuum state are independent of $\xi$. In principle one could also consider a massive theory, however for such a quantum field the past asymptotic vacuum state has a more complicated form (see Sec.\ \ref{sec:massive}), and hence we shall not consider that state here.} Such a space-time is described by the ingoing Vaidya metric
	\begin{equation}
	\label{eq:vaidya}
	\diff s^2=-\left(1-\frac{\vartheta(t)R_s}{r}\right)\diff v^2+2dRdv+R^2\diff \Omega^2,
	\end{equation}
	where $\vartheta$ is the Heaviside step function (see Ref.\ \cite{poissonRelativistToolkitMathematics2004}).\footnote{This set-up is also considered in Ref.\ \cite{andersonMethodComputeStressenergy2020}, where a mode-sum approach to evaluation of the stress-tensor is proposed.} We would like to bring this metric in quasi-conformal form (i.e. in the form of eq.\ \eqref{eq:metric}), which are in particular outgoing coordinates. Typically, it may not be possible to cover the full space-time with a quasi-conformal globally hyperbolic coordinate patch, but in this case we find two distinct patches that lend themselves for answering different questions regarding the Hawking effect. Patch I covers the entire black hole exterior, and the coordinates are uniquely fixed by matching them to outgoing Vaidya coordinates in the post collapse region. This patch is well-suited to re-derive the original results due to Hawking for this collapse model, namely the radiation spectrum at future asymptotic infinity. Patch II covers a neighbourhood of the collapsing shell, inculding the path of the shell inside the black hole interior, and is therefore more suited to gather information about non-linear observables during gravitational collapse.
	
	\subsection{Constructing quasi-conformal coordinates}
\begin{figure}
		\centering
		\begin{tikzpicture}[scale=.8]
		\fill[blue!10] (0,0)--(4,4)--(2.5,5.5)--(0,3);
		\draw (0,0)--(4,4)--(2.5,5.5);
		\draw[dashed] (0,0)--(0,5.5);
		\draw[zigzag] (0,5.5)--(2.5,5.5);
		\draw (0.1,-.2) node {$i^-$} (2.25,1.75) node {$\mathscr{I}^-$} (4.2,4) node {$i^0$} (3.5,5) node {$\mathscr{I}^+$} (2.6,5.75) node {$i^+$};
		\end{tikzpicture}
		\caption{A Penrose diagram of the space-time with the Vaidya metric of eq.\ \eqref{eq:vaidya} with the domain of coordinate patch I as shaded region}
		\label{fig:patch1}
	\end{figure}
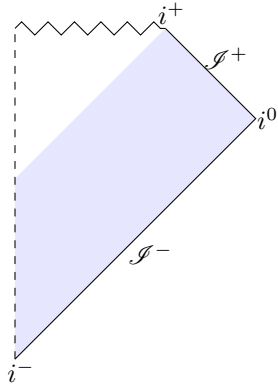

	The first step in constructing the quasi-conformal coordinates, is to go to double-null coordinates. This entails solving the equation
	\begin{equation}
	\label{eq:Rpdv}
	\partial_vR(v,u)=\frac{1}{2}\left(1-\frac{\theta(v)R_s}{R(v,u)}\right),
	\end{equation} where $R$ is the radial coordinate of the Vaidya space-time, such that $4\pi R^2$ measures the area of a 2-sphere centered around the axis of symmetry. For patch I, we demand that 
	\begin{equation}
		\lim_{v\rightarrow\infty}\frac{2R_I(v,u)}{v-u}=1.
	\end{equation}
	This gives a solution
	\begin{align}
	R_I(v,u)=&\vartheta(v)\left(R_s+R_sW\left(\exp\left(\frac{v-u}{2R_S}-1\right)\right)\right)\nonumber\\
	&+\vartheta(-v)\left(\frac{v}{2}+R_s+R_sW\left(\exp\left(\frac{-u}{2R_S}-1\right)\right)\right).
	\end{align}
	Here $W$ is the prime branch of the Lambert W-function (see Ref.\ \cite{vebericLambertFunctionApplications2012}) and $u$ has as domain the full real number line. The metric now takes the form
	\begin{equation}
		\diff s^2=2\pdv{R(v,u)}{u}\diff u\diff v+R(v,u)^2\diff\Omega^2.
	\end{equation}
	To transform to quasi-conformal coordinates, we leave $u$ unchanged, but define the coordinate $r$ by the relation
	\begin{equation}
		\pdv{r_I^{-1}}{v}=-\pdv{R_I^{-1}}{u}.
	\end{equation}
	For this choice of coordinates, a solution to the relation above, with boundary condition 
	\begin{equation}
		\lim_{v\rightarrow\infty}r_I(v,u)^{-1}=0,
	\end{equation} yields a well defined (non-negative) function that is strictly increasing in $v$ for the full black hole exterior (and hence one can inverse this function to get a well defined $r$ coordinate on the full exterior). This means we can define the quasi-conformal coordinate patch I on the full black hole exterior, as drawn in Figure \ref{fig:patch1}. In particular, one finds
	\begin{equation}
		r_I(v,u)=\vartheta(v)R_I(v,u)+\vartheta(-v)\frac{R_I(0,u)^2(v+2R_I(0,u))}{2R_I(0,u)^2+R_Sv}.
	\end{equation}
	Using $(u,r,\Omega)$ coordinates, the metric takes quasi-conformal form, with
	\begin{equation}
	\exp(2\alpha_I(u,r))=\vartheta(r-R_I(0,u))\left(1-\frac{R_S}{r}\right)+\vartheta(R_I(0,u)-r)\frac{R_I(0,u)^3 
		-3rR_I(0,u)R_S + 
		2 r^2R_S}{R_I(0,u)^3},
	\end{equation}
	and
	\begin{equation}
	\exp(\beta_I(u,r))=1-\vartheta(R_I(0,u)-r)R_S\frac{R_I(0,u)-r}{R_I(0,u)^2-rR_S}.
	\end{equation}
	Using the quasi-conformal metric defined by functions $\alpha_I$ and $\beta_I$ as above to calculate $\kappa_u^{(0)}$ from equation \eqref{eq:kappa0}, one finds that for $r<R_I(0,u)<r'$
	\begin{equation}
		\kappa_{u,I}^{(0)}(r,r')=R_S\frac{ (r+r')(R_I(0,u)^3-3rr'R_I(0,u))+4r^2r'^2}{4  r r' (r'-r)^3
		R_I(0,u)^3},
	\end{equation}
	and that it is zero for $r,r'<R_I(0,u)$ and $r,r'>R_I(0,u)$.
	This function plays a role in the source term for $\tilde{\lambda}_u$ (as defined in \eqref{eq:char_had_ans}), in particular
	\begin{equation}
		(\partial_t\partial_r\partial_{r'}-K_t\partial_{r'}-K'_t\partial_r)RR'\tilde{\lambda}_u(r,\Omega;r',\Omega')=-2\pi\kappa_u^{(0)}(r,r')\delta(\Omega,\Omega').
	\end{equation}
	This relation is admittedly somewhat formal, as it only really holds when both sides of the equations are smeared with $\tilde{S}_{s.c.}(C_u)$ functions (as defined in Sec.\ \ref{sec:qft_bound_setup}).\footnote{Note in particular that the asymptotic mode functions typically used to calculate the spectrum of the radiation produced in gravitational collapse, as observed by an observer near $\mathscr{I}^+$, cannot be embedded into $\tilde{S}_{s.c.}(C_u)$ via ($R$-weighted) restriction to $C_u$. Not only because these mode functions are not spatially compact, but also because mode function typically diverge near the tip of $C_u$. This means in particular that the relation above cannot be expected to hold when smeared with mode-functions, in fact the right hand side will typically be divergent. Therefore, one cannot use this relation directly to calculate the spectrum of the produced radiation. This can be mended by going to a higher order of Hadamard subtraction, for which the resulting source term (given by eq.\ \eqref{eq:source} cut off at finite $n$) is more well-behaved near $r=0$ and hence can be integrated with a larger class of functions.} Nevertheless we can loosely think of $\kappa_u^{(0)}$ as sourcing the `particle production' for our quantum field living on a gravitational collapse space-time. This really highlights the fact that the produced radiation originates near the collapsing body, as opposed to near the horizon, as is sometimes suggested.\footnote{It is slightly suggestive to speak of a region of origin of this radiation in the first place. As can be seen from the support of $\kappa_u^{(0)}$, the correlations of the two-point functions between past and post collapse regions (i.e. the regions separated by $r=R_I(0,u)$) are what is primarily sourced by $\kappa_u^{(0)}$. Of course these correlations then propagate to asymptotic infinity, where they then read as radiation coming from the direction of the collapsing body.} As discussed in detail in \cite{fredenhagenDerivationHawkingRadiation1990}, the presence of a Killing horizon in the post collapse region implies that the radiation observed at $\mathscr{I}^+$ asymptotes towards a thermal spectrum, simply due to the universal nature of a black hole geometry near such an horizon. Verifying the thermality of the radiation produced in this particular collapse process, requires calculating the asymptotic spectrum, which is a calculation we shall leave for another time. What we will note is that the fact that the observed radiation asymptotes to a steady state (such as thermal radiation), can already be seen directly from $\kappa_u^{(0)}$, as for $r<R_{I}(0,u)<r'$ this function asymptotes to
	\begin{equation}
		\lim_{u\rightarrow\infty}\kappa_u^{(0)}(r,r')=\frac{ (r+r')(R_S^3-3rr'R_S)+4r^2r'^2}{4  r r' (r'-r)^3
		R_S^2}.
	\end{equation}

\begin{figure}
		\centering
		\begin{tikzpicture}[scale=.8]
		\fill[red!10] (0,0)--(4,4) to [out=225,in=315] (0,5.5)--(0,0);
		\draw[dotted] (0,5.5)--(2.75,2.75);
		\draw (0,0)--(4,4)--(2.5,5.5);
		\draw[dashed] (0,0)--(0,5.5);
		\draw[zigzag] (0,5.5)--(2.5,5.5);
		\draw (0.1,-.2) node {$i^-$} (2.25,1.75) node {$\mathscr{I}^-$} (4.2,4) node {$i^0$} (3.5,5) node {$\mathscr{I}^+$} (2.6,5.75) node {$i^+$};
		\end{tikzpicture}
		\caption{A Penrose diagram of the space-time with the Vaidya metric of eq.\ \eqref{eq:vaidya} with the domain of coordinate patch II as shaded region, and the dotted line denoting the collapsing shell}
		\label{fig:patch2}
	\end{figure}
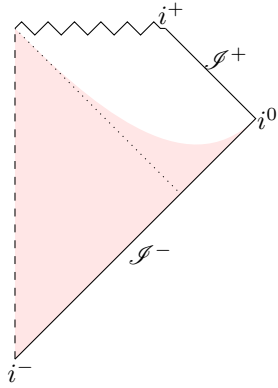

	Shifting away from patch I, we will focus with more detail on coordinate patch II. We construct this patch by solving $\eqref{eq:Rpdv}$, where we demand that for $v<0$
	\begin{equation}
		R_{II}(v,u)=\frac{v-u}{2}.
	\end{equation}
	This gives for $v>0$ that
	\begin{equation}
		R_{II}(v,u)=R_S+R_SW\left(-\frac{u+2R_S}{2RS}\exp(\frac{v-u-2R_S}{2R_S})\right).
	\end{equation}
	Here the domain of $u$ is $(-\infty,0)$. Similarly to the null coordinates used to construct patch I, the function $R_{II}$ above can be used to construct null coordinates on the black hole geometry. However, in this instance these null coordinates cover the full space-time, where the metric again takes the form
	\begin{equation}
		\diff s^2=2\pdv{R_{II}(v,u)}{u}\diff u\diff v+R_{II}(v,u)^2\diff \Omega^2.
	\end{equation}
	Using once more the relation
	\begin{equation}
		\pdv{r_{II}^{-1}}{v}=-\pdv{R_{II}^{-1}}{u},
	\end{equation}
	to define quasi-conformal coordinates $(u,r_{II},\Omega)$, we find that for $u\neq-2R_S$
	\begin{equation}
		r_{II}(v,u)=\vartheta(-v)\frac{v-u}{2}+\vartheta(v)\frac{-uR_{II}(v,u)(u+2R_S)}{4R_SR_{II}(v,u)-u^2},
	\end{equation}
	while for $u=-2R_S$ this relation can be extended by continuity to
	\begin{equation}
		r_{II}(v,-2R_S)=\vartheta(-v)\frac{v+2R_S}{2}+\vartheta(v)\frac{R_S}{2-\exp(\frac{v}{2R_S})}.
	\end{equation}
	It should be noted that $r_{II}$ blows up near the surface 
	\begin{equation}
		R_{II}(v,u)=\frac{u^2}{4R_S},
	\end{equation}
	hence these quasi-conformal coordinates can only be used to cover the black hole space-time up to this boundary.\footnote{Strictly speaking this relation defines this surface for $u\neq-2R_S$, by continuity this relation can then be extended to $u=-2R_S$, i.e. where the surface crosses the horizon, which happens at $v=2R_S\ln(2)$.} Luckily, the full trajectory of the collapsing body is contained within the patch where $r_{II}$ is defined. Furthermore, we can calculate the induced metric on this boundary surface, namely
	\begin{equation}
		\diff s^2=\left(1-\frac{u}{R_S}\right)\diff u^2+\frac{u^4}{16R_s^2}\diff \Omega^2,
	\end{equation}
	which in particular means that this (future) boundary is space-like. Therefore, patch II, drawn in Figure \ref{fig:patch2}, is globally hyperbolic.
	
	On the quasi-conformal coordinate patch II, we can now calculate the metric coefficients to be
	\begin{equation}
	\exp(2\alpha_{II})=1-\vartheta(2r+u)R_S(2r+u)^2\frac{2R_S(2r+u)-(4r-u)u}{ru^4},
	\end{equation}
	\begin{equation}
	\exp(\beta_{II})=1-\vartheta(2r+u)\frac{2R_S(2r+u)}{\left(u^2+2R_S(2r+u)\right)}.
	\end{equation}
	\subsection{Calculating Wick squares on a Vaidya space-time}
	Now that we have found a quasi-conformal coordinate patch surrounding the collapsing shell, we want to give an expression for expectation values locally covariant non-linear observables that can be used for explicit calculations. As a proof of concept, we shall restrict our attention to $:\Phi^2:$. We consider the two-point function in the asymptotic past vacuum state. Recalling the discussion of Sec.\ \ref{sec:renorm}, for a two-point function restricted to $C_u$ of the form 
	\begin{align}
	\Lambda(u,r,\Omega;u,r',\Omega')=\frac{1}{8\pi^2}\Bigg[&\frac{\exp(-\beta-\beta')}{rr'(1-\cos\theta)+\textup{i}0^+(r-r'-i0^+)}\nonumber\\&+v_u^{(0)}(r,r')\ln(RR'(1-\cos\theta))+w_u^{(0)}(r,\Omega;r',\Omega')\Bigg],
	\end{align}
	with $w_u^{(0)}(r,\Omega;r',\Omega')$ a $C^1$ function, the expectation value of $:\Phi^2:$ is given by the expression
	\begin{equation}
		\langle:\Phi^2:\rangle_{\Lambda}=\frac{1}{8\pi^2}w_u^{(0)}(r,\Omega;r,\Omega)+Q_{\Phi^2}.
	\end{equation}
	The covariance correction term can be easily calculated to be
	\begin{equation}
		Q_{\Phi^2}=\vartheta(2r+u)\frac{R_S(u^2+2R_S(2r+u))^3}{48\pi^2r^3u^6}.
	\end{equation}
	Calculating $w_u^{(0)}(r,\Omega;r,\Omega)$ is a bit more involved. First we establish that for $r<\frac{-u}{2}<r'$
	\begin{multline}\kappa_{u,II}^{(0)}(r,r')=\\R_s\frac{u\left(u^3(r+r')-12urr'(r+r')-36r^2{r'}^3\right)+2R_S(2r+u)(2r'+u)(r(2r'+u)+r'(2r+u))}{rr'u^4(r'-r)^3},
	\end{multline}
	and that $\kappa_{u,II}^{(0)}(r,r')=0$ for $r,r'<-\frac{u}{2}$ and $r,r'>-\frac{u}{2}$. Using relation \eqref{eq:lambda_recursion}, we find
	for $r<\frac{-u}{2}<r'$ that
	\begin{multline}
	v_{u,II}^{(0)}(r,r')=-\frac{R_S(2r+u)(2r'+u)(u^2+2R_S(2r'+u))}{4u^7(r'-r)^3}\\\times\left(u(u^2-2(r+r')u-12rr')+2R_s(2r+u)(2r'+u)\right),
	\end{multline}
	and also that this function is zero for $r,r'<-\frac{u}{2}$ and $r,r'>-\frac{u}{2}$.
	We can now write a dynamical equation for $w_u^{(0)}(r,\Omega;r',\Omega')$, namely
	
	\begin{align}
	\left(\partial_u\partial_r\partial_{r'}+K_u\partial_r+K'_u\partial_{r'}\right)RR'w_u^{(0)}=&-\ln(RR'(1-\cos\theta))\kappa_u^{(1)}\nonumber\\
	&-\frac{1}{2}\Big[\frac{1}{RR'}\partial_uRR'\partial_{r}\partial_{r'}RR'v_u^{(0)}\nonumber\\&+\partial_u\left((\partial_rR)(\partial_{r'}R')v_u^{(0)}-RR'\partial_r\partial_{r'}v_u^{(0)}\right)\nonumber\\
	&+\frac{R'}{r'^2}\partial_{r}Rv_u^{(0)}\nonumber\\
	&+(\partial_{r'}R')\left(\exp(2\beta)\tilde{V}-\partial_r\exp(2\alpha)\partial_r\right)Rv_u^{(0)}\nonumber\\
	&-\frac{1}{R}(\partial_rR)\exp(2\alpha)\partial_r\partial_{r'}RR'v_u^{(0)})\nonumber\\
	&-\partial_r\exp(2\alpha)(\partial_rR)\partial_{r'}R'v_u^{(0)})\nonumber\\
	&+(r\leftrightarrow r')\Big]=:S_u^{(0)}(r,\Omega;r',\Omega').
	\end{align}
	Using the formal integration kernel of the map $\rho_u:\mathcal{D}(M_u)\rightarrow\tilde{S}_{s.c.}(C_u)$, given by $\rho_u(r,\Omega;\overline{u},\overline{r},\overline{\Omega})$, we can write down a solution to this equation, provided the initial data $\lim_{u\downarrow-\infty}RR'w_u^{(0)}(r,r')=0$. That being
	\begin{align}
	w_{u_0}^{(0)}(r_0,\Omega_0;r_0',\Omega_0')=4&\int_{-\infty}^{u_0}\diff u\int \diff r\int \diff\Omega\int \diff r'\int \diff\Omega'\nonumber\\&\rho_u(r,\Omega;u_0,r_0,\Omega_0)\rho_u(r',\Omega';u_0,r_0',\Omega_0')S_{u}^{(0)}(r,\Omega;r',\Omega').
	\end{align}
	While we cannot bring this expression into closed form, we point out that this equation may be used as the starting point for a numerical calculation of $w_u$.\footnote{Admittedly, numerical calculations involving retarded propagators on black hole space-times are far from easy to implement, especially as they themselves have non-trivial singular behaviour. See for instance Ref.\ \cite{casalsRegularizedCalculationRetarded2019} where this problem is studied on a Schwarzschild background.} Alternatively, we may expand this equation in powers of $R_S$. Clearly in the limit $R_S\rightarrow 0$, coordinate patch $II$ reduces to (a patch of) Minkowski space-time. Writing the operator $K_t=K_t^{(0)}+R_SK_t^{(1)}$, with $K_t^{(0)}$ the operator in the Minkowski limit (i.e. for $R_S=0$) we can write $w_u^{(0)}(r,\Omega;r',\Omega')$ as a formal series
	\begin{equation}
		w_u^{(0)}(r,\Omega;r',\Omega')=\sum_{n=0}^\infty R_S^nw_{u,n}^{(0)}(r,\Omega;r',\Omega'),
	\end{equation}
	where 
	\begin{multline}
	w_{u_0,0}^{(0)}(r_0,\Omega_0;r_0',\Omega_0')=4\exp(-\beta(u_0,r_0)-\beta(u_0,r_0'))\int_{-\infty}^{u_0}\diff u\int \diff r\int \diff \Omega\int \diff r'\int \diff \Omega'\\\rho_{u,0}(r,\Omega;u_0,r_0,\Omega_0)\rho_{u,0}(r',\Omega';u_0,r_0',\Omega_0')S_{u}^{(0)}(r,\Omega;r',\Omega'),
	\end{multline}
	with $\rho_{u,0}$ the ($r$-weighted) causal propagator on Minkowski space-time, and 
	\begin{multline}
	w_{u_0,n+1}^{(0)}(r_0,\Omega_0;r_0',\Omega_0')=-4\exp(-\beta(u_0,r_0)-\beta(u_0,r_0'))\int_{-\infty}^{u_0}\diff u\int \diff r\int \diff \Omega\int \diff r'\int \diff \Omega'\\\rho_{u,0}(r,\Omega;u_0,r_0,\Omega_0)\rho_{u,0}(r',\Omega';u_0,r_0',\Omega_0')\times\\(K^{(1)}_u\partial_{r'}+{K^{(1)}_u}'\partial_r)R(u,r)R(u,r')w_{u,n}^{(0)}(r,\Omega;r',\Omega'),
	\end{multline}
	The terms in this series expansion may be calculated numerically, but in principle they may also be calculated by hand. Here we show how to calculate $w_u(r,\Omega;r,\Omega)$ to lowest order in $R_S$. We make two observations, firstly that for $u_0>u$ we have that
	\begin{align}
	\rho_{u,0}(r,\Omega;u_0,r_0,\Omega_0)=&\frac{r\delta\left(u_0-u+r_0-r-\sqrt{r^2+r_0^2-2rr_0\cos\theta}\right)}{4\pi(u_0-u+r_0-r)}\nonumber\\=&\frac{\delta\left(\cos\theta-1+\frac{(u_0-u)(u_0-u+2(r_0-r))}{2rr_0}\right)}{4\pi r_0}.
	\end{align}
	Secondly, to calculate the contribution due to the logarithmic term in $S_u^{(0)}$, we shall use that for $\vert x\vert,\vert x'\vert<1$, we can calculate the integral
	\begin{equation}
		\int_0^{2\pi} \diff \phi\ln(1-xx'-\cos\phi\sqrt{1-x^2}\sqrt{1-{x'}^2})=2\pi\ln(\frac{1-xx'+\vert x-x'\vert}{2}).
	\end{equation}
	This can be verified most easily by expanding the integrand in power series of $\cos\theta$. 
	We can now explicitly work out the coincidence limit of $w_u$ to lowest order in $R_S$, namely
	\begin{align}
	w_{u_0}^{(0)}(r_0,\Omega_0;r_0,\Omega_0)=&4\int_{-\infty}^{u_0}\diff u\int \diff r\int \diff \Omega\int \diff r'\int \diff \Omega'&\rho_{u,0}(r,\Omega;u_0,r_0,\Omega_0)\rho_{u,0}(r',\Omega';u_0,r_0,\Omega_0)\nonumber\\
	&&\times S_{u}^{(0)}(r,\Omega;r',\Omega')+\mathcal{O}(R_S^2)\nonumber\\=&-\vartheta(2r_0+u_0)\frac{R_S}{6r_0^3}+\mathcal{O}(R_S^2),
	\end{align}
	We find that this contribution to $\langle:\Phi^2:\rangle$ exactly cancels that of the covariance correction term $Q_{\Phi}$, and hence we find
	\begin{equation}
	\langle:\Phi^2:\rangle(u,r)=\mathcal{O}(R_S^2).
	\end{equation}
	At this point we do not know if there is a particularly deep reason for this lowest order term to vanish. In similar fashion as what is done in Ref.\ \cite{juarez-aubryQuantumFieldsBlack2018} for the 1+1 dimensional case, one would like to compare this result to recent numerical computations of $\langle:\Phi^2:\rangle$ in the Unruh state, such as performed in Ref.\ \cite{lanirModesumRenormalizationMathrm2018}. However, to properly make such a comparison, such first order calculations as performed above are insufficient. At this point we do not know the rate of convergence of the series expansions for any non-linear observable calculated in this way. Hence, the first order term need by no means be the most dominant, especially not once one evaluates these quantities near or beyond the event horizon. Here it should be especially noted that in our method the expansions are given at fixed dimensionfull coordinates, while if one for instance wants to evaluate $\langle:\Phi^2:\rangle$ at the Schwarzschild radius, the radial coordinate would also need to be rescaled with respect to $R_S$. That being said, the functions $\alpha$ and $\beta$, from which these expectation values are ultimately calculated, have expansions in $R_S$ that for a fixed value of this expansion parameter have better convergence rates as one approaches the collapsing shell. Therefore we suspect that a similar fact may hold for the expansions of $\langle:\Phi^2:\rangle$.
	
	Of course the mere expecation value of $\Phi^2$ does not tell us much about the estimated back-reaction of the field onto the geometry, for which we would at least need to calculate the expectation value of the stress-energy tensor. Still, generalizing these methods to calculate the relevant components of the stress-energy tensor do not require so much extra effort. However, for these observables the thin shell collapse model used in this section may not be the most practical, as given the low regularity of this space-time one may expect these observables to diverge near the collapsing shell. Thus, in future work we hope to use the methods developed above to calculate the stress-energy tensor of a quantum field on more realistic (and regular) gravitational collapse space-times, both analytically as well as numerically. We hope that this will give us deeper insight into the back-reaction of quantum fields onto the background geometry during gravitational collapse.
	
	\section{Concluding remarks}
	As far as the authors are concerned, the take-home messages of this paper are the following: for spherically symmetric space-times, the singular part of the Hadamard two-point functions restricted to null cones can be described entirely in terms of particular coordinates adapted to the light-cone and in terms of derivatives of metric components at that cone. Here the amount of time-derivatives (w.r.t. a time-function that has the null cone as a level surface) necessary to write down a characteristic Hadamard parametrix modulo $C^{2N+1}$ is at most $N+1$, where for conformal coupled field this number even reduces to $N$. While the overall form of the singular structure is by no means surprising, as it can be viewed as a restriction of the usual Hadamard series to the characteristic cone, the fact that this characteristic parametrix can be naturally constructed on the full null cone (provided it can be covered by quasi-conformal coordinates), gives us some novel tools in studying the dynamics of Hadamard states, and how they backreact onto a dynamical background geometry. In particular, we have seen how the characteristic Hadamard parametrix can be used to write down a renormalization scheme for non-linear observables, where the separation between the state-dependent and universal part of the two-point function is made such that, at least for conformally coupled theories, the semi-classical Einstein equations somewhat simplify to a pair of differential equations that are first order in time. Whether this actually means that the semi-classical Einstein equations have solutions for any interesting sets of initial data on some null cone, i.e. whether their Goursat problem is well posed, is a question we hope to answer in the future. For now we will end on the point that the use of the characteristic Hadamard parametrix as a tool to locally covariantly renormalize non-linear observables gives us a way to explicitly calculate the expectation value of such observables as some series expansion. It is hoped that these calculations can also be numerically implemented, yielding a calculation method for observables of quantum fields that, as opposed to more involved methods of calculating locally covariant renormalized observables using numerics, does not involve mode expansions of Hadamard states and Hadamard parametrices.
	
	\section{Acknowledgements}
	We would like to thank Nicola Pinamonti and Paolo Meda for fruitful discussions, the former also for hosting D.W.\ Janssen during his research visit to the Dipartimento di Matematica at the University of Genova. The work of D.W.\ Janssen is
	funded by the Deutsche Forschungsgemeinschaft (DFG) under Grant No.\ 406116891 within
	the Research Training Group RTG 2522: ``Dynamics and Criticality in Quantum and Gravitational Systems" (\texttt{http://www.rtg2522.uni-jena.de/}).
	\newpage
	\appendix
	\section{Definitions from microlocal anlysis}
	\label{app:micro}
	In this paper we give recite characterizations of (quasi-free) Hadamard states in terms of their wave front sets. These are objects associated with a distribution that play a central role in microlocal analysis. Here we shall recall the relevant definitions. For a more complete and contextualized overview, see Ref.\ \cite{hormanderAnalysisLinearPartial2015}.
	
	We shall first define the notion of a distribution.
	\begin{definition}
		On an open subset $V\subset \mathbb{R}^n$, the set of \textup{distributions} $\mathcal{D}'(V)$ are linear maps $u:\mathcal{D}(V)\rightarrow \mathbb{R}$, where $\mathcal{D}(V)$ (real) vector space of compactly supported real-valued smooth functions on $V$, such that for each compact $K\subset V$ there exists a $C\in\mathbb{R}_{>0}$ and an $N\in\mathbb{N}$ such that
		\begin{equation}
		\vert u(f)\vert\leq C\sup\{\vert\partial_{x_1}^{\alpha_1}...\partial_{x_n}^{\alpha_n}f(x)\vert:x\in K,\,\alpha_i\in\mathbb{N},\,\sum_{i=1}^n\alpha_i\leq N\}.
		\end{equation}
		
		On an smooth $n$-manifold $M$, which has an atlas $(U_i,\phi_i)_{i\in I}$ consisting of an open cover $U_i\subset M$ and homeomorphisms $\phi_i:U_i\rightarrow V_i\subset\mathbb{R}^n$ such that for each $i,j\in I$ the map $\phi_i\circ\phi_{i'}^{-1}$ is smooth,
		a \textup{distribution density} $\tilde{u}:\mathcal{D}(M)\rightarrow \mathbb{R}$ is a linear map such that for each $i\in I$ one can define a distribution $u_i\in\mathcal{D}'(V_i)$ via
		\begin{equation}
			u_i(f)=\tilde{u}(f\circ\phi_i),
		\end{equation}
		with $f\circ\phi_i\in\mathcal{D}(M)$ defined to be 0 outside $U_i$.
	
		On a space-time $M$ with metric $g$, the set of \textup{distributions} $\mathcal{D}'(M)$ are the linear maps $u:\mathcal{D}(M)\rightarrow\mathbb{R}$ such that $\tilde{u}=\diff\textup{vol}_gu$ is a distribution density. This means that for $u_i$ defined above we now have
		\begin{equation}
			u_i(f)=u\left(\left(f\sqrt{\vert g\vert}\right)\circ\phi_i\right),
		\end{equation}
		with $\vert g\vert$ the absolute value of the determinant of the metric $g$ in local coordinates defined by $\phi_i$.\footnote{One can define distributions on a manifold without the use of a smooth density, see Ref.\ \cite[Def.\ 6.3.3.]{hormanderAnalysisLinearPartial2015}. In a setting with a canonical choice of strictly positive smooth density defining the integration measure on a manifold, such as in the case of a Lorentzian space-time, distributions and distribution densities are canonically identified with one another as in the definition above.}
		
		The space of distributions of compact support, i.e. maps $u\in\mathcal{D}'(M)$ for which there is a compact $K\subset M$ such that for each $f\in\mathcal{D}(M)$ with $\textup{supp}(f)\cap K=\emptyset$ we have $u(f)=0$ (where the smallest $K$ for which this holds defines $\textup{supp}(u)$), are denoted by $\mathcal{E}'(M)$. $u\in\mathcal{E}'(M)$ can be viewed as a map $u:\mathcal{E}(M)\rightarrow \mathbb{R}$ via the extension given by
		\begin{equation}
			(hu)(f):=u(hf),
		\end{equation}
		with $h\in\mathcal{D}(M)$ and $h(x)=1$ for each $x\in\textup{supp}(u)$.
	\end{definition}
	To define the wave front set, we introduce the Fourier transform of a compactly supported distribution on $\mathbb{R}^n$.
	\begin{definition}
		Given $u\in\mathcal{E}(\mathbb{R}^n)$, we can define the \textup{Fourier} transform $\hat{u}\in\mathcal{E}(\mathbb{R}^n,\mathbb{C})$ via
		\begin{equation}
			\hat{u}(k)=u(\cos(\langle.,k\rangle)-\textup{i} u(\sin(\langle.,k\rangle)),
		\end{equation}
		where $\langle.,.\rangle$ is the standard Euclidean inner product on $\mathbb{R}^n$. 
	\end{definition}
	That this function is indeed smooth follows from Ref.\ \cite[Thm.\ 2.1.3.]{hormanderAnalysisLinearPartial2015}. We now define the wave front set of a distribution.
	\begin{definition}
		For $V\subset\mathbb{R}^n$ open and $u\in\mathcal{D}(V)$, the \textup{wave front set} $WF(u)\subset V\times(\mathbb{R}^n\setminus \{(0,...,0)\})\cong T^*V\setminus\mathbf{0}$ is defined by $(x,k)\not\in WF(u)$ if and only if there is an $h\in\mathcal{D}(V)$ with $h(x)\neq 0$ and a conic open neighbourhood $k\in\Gamma\subset \mathbb{R}^n$ such that for each $N\in\mathbb{N}_{>0}$ there is a $C>0$ with for any $\zeta\in \Gamma$
		\begin{equation}
			\vert \hat{(hu)}(\zeta)\vert\leq C(1+\vert\zeta\vert)^{-N}.
		\end{equation}
		
		Let $M$ a space-time of dimension $n$ with atlas $(U_i,\phi_i)_{i\in I}$ and let $u\in\mathcal{D}'(M)$. The set $WF(u)\subset T^*M\setminus\mathbf{0}$, with $T^*M$ the cotangent bundle and $\mathbf{0}\cong M\times \{(0,...,0)\}$ the zero bundle, is given by $(x,k)\in WF(u)$ iff there is an $i\in I$ with $(x,k)\in T^*U_i$ such that $(\phi_i(x),\left(\phi_i^{-1}\right)^*(k))\in WF(u_i)$.		
	\end{definition}
	The wave front set of a distribution tells us what operations defined on smooth functions (multiplications, restrictions etc.) can be naturally extended to it (see \cite[Ch.\ 8.2.]{hormanderAnalysisLinearPartial2015}). Furthermore, the wave front set plays a crucial role in the propagation of singularities of weak solutions to differential equations (see Ref.\ \cite[Sec.\ 6.1]{duistermaatFourierIntegralOperators1972}).

	In this paper we mainly use the wave front set in the context of two-point functions. As described in Sec.\ \ref{sec:qftsetup}, for a real scalar field on a globally hyperbolic space-time $M$, a two-point function is given by a map $\Lambda:\mathcal{D}(M)^2\rightarrow\mathbb{C}$ with
	\begin{equation}
		\Lambda(f,g)=\mu(f\otimes g)+\frac{\textup{i}}{2}E(f,g),
	\end{equation}
	where $\mu\in\mathcal{D}'(M\times M)$ with in particular $\mu(f\otimes f)\geq 0$, $\mu(f\otimes Pg)=\mu(Pf\otimes g)=0$ and $E(f,g)^2\leq4\mu(f,f)\mu(g,g)$. For such bi-distributions, which uniquely extend to distributions on $M\times M$, one often works with the primed wave front set $WF'(\Lambda)\subset (T^*M)^2\setminus\mathbf{0}^2$, which is defined as
	\begin{equation}
		(x_1,k_1;x_2,k_2)\in WF'(\Lambda)\iff ((x_1,x_2),(k_1,-k_2))\in WF(\Lambda)\subset T^*(M\times M)\setminus\mathbf{0}.
	\end{equation}
	For example, for the commutator function $E$ given by Eq. \eqref{eq:comm}, one has
	\begin{equation}
		WF'(E)=\{(x_1,k_1;x_2,k_2)\in(T^*M\setminus\mathbf{0})^2:g^{\mu\nu}(x_i)(k_i)_\mu(k_i)_\nu=0,\;(x_1,k_1)\sim(x_2,k_2)\}
	\end{equation}
	where the equivalence relation $\sim$ on null covectors is defined as $(x_1,k_1)\sim(x_2,k_2)$ if there exists a null geodesic $\gamma:\mathbb{R}\rightarrow M$ with $a,b\in\mathbb{R}$ such that \begin{equation}
		\gamma(a)=x_1,\, g_{\mu\nu}(x_1)\dot\gamma(a)^\mu=(k_1)_\nu,\, \gamma(b)=x_2,\, g_{\mu\nu}(x_2)\dot\gamma(b)^\mu=(k_2)_\nu.
	\end{equation}
	A relevant operation on (primed) wave-front sets of bi-distributions is the following.
	\begin{definition}
		Given $\Lambda$ a bi-distribution on $M$,or equivalently $\Lambda\in\mathcal{D}'(M\times M)$, we define
		\begin{equation}
			{}_{M}WF'(\Lambda)=\{(x_2,k_2)\in T^*M\setminus\mathbf{0}:(x_1,0;x_2,k_2)\in WF'(\Lambda)\text{ for some }x_1\in M\},
		\end{equation}
		and
		\begin{equation}
		WF'(\Lambda)_{M}=\{(x_1,k_1)\in T^*M\setminus\mathbf{0}:(x_1,k_1;x_2,0)\in WF'(\Lambda)\text{ for some }x_2\in M\}.
		\end{equation}
	\end{definition}
	These objects correspond to the wave front sets of $\Lambda$ smeared with a single $f\in\mathcal{D}(M)$. In particular, as proven in Ref.\ \cite[Thm.\ 8.2.12.]{hormanderAnalysisLinearPartial2015},
	\begin{equation}
		WF(\Lambda(f,.))=-{}_{M}WF'(\Lambda),\, WF(\Lambda(.,f))=WF'(\Lambda)_{M}.
	\end{equation}
	\newpage
	\section{Detailed derivations of formulas}
	\label{app:dervs}
	Here we write out the derivations of certain formulas from the main text in some more detail. 
	\subsection{Equation \eqref{eq:char_ker_dyn}}
	This relation is somewhat formal, meaning that it may not be well-defined for all distributions, hence one has to verify whether this relation holds for a particular class of distributions at hand. Let us in this case consider a family of functions $\lambda_t\in L_1(C_t^2,\diff r\diff r'\diff \Omega\diff\Omega')$ for $t\in\mathbb{R}$ such that $\partial_t\lambda_t\in L_1(C_t^2,\diff r\diff r'\diff \Omega\diff\Omega')$. This family generates bi-distributions $\Lambda_t\in\mathcal{D}(M_t\times M_t)$ in the bulk via
	\begin{equation}
		\Lambda_t(f,g)=\int_0^\infty\diff r\int_0^\infty\diff r'\int_{S^2}\diff\Omega\int_{S^2}\diff\Omega'\lambda_t(r,\Omega;r',\Omega')\rho_t(f)(r,\Omega)\rho_t(g)(r',\Omega').
	\end{equation}
	We assume that for given $f,g\in\mathcal{D}(M_t)$ there is a $\delta>0$ such that for $t-\delta<t'<t+\delta$ we have $\Lambda_t(f,g)=\Lambda_{t'}(f,g)$, which means in particular that
	\begin{equation}
		\dv{\Lambda_t(f,g)}{t}=0.
	\end{equation}
	Let us furthermore assume for each $t'\in(t-\delta,t+\delta)$ there exists a $\tilde{\lambda}\in L_1(C_t^2,\diff r\diff r'\diff \Omega\diff\Omega')$ such that $\vert\lambda_{t'}\vert\leq\tilde{\lambda}$ almost everywhere on $C_t^2$. In this case we can use the Leibniz integral rule to derive
	\begin{align}
		0=&\dv{t}\int_0^\infty\diff r\int_0^\infty\diff r'\int_{S^2}\diff\Omega\int_{S^2}\diff\Omega'\,\lambda_t(r,\Omega;r',\Omega')\rho_t(f)(r,\Omega)\rho_t(g)(r',\Omega')\nonumber\\
		=&\int_0^\infty\diff r\int_0^\infty\diff r'\int_{S^2}\diff\Omega\int_{S^2}\diff\Omega'\,\left(\partial_t\lambda_t(r,\Omega;r',\Omega')\right)\rho_t(f)(r,\Omega)\rho_t(g)(r',\Omega')\nonumber\\
		&+\int_0^\infty\diff r\int_0^\infty\diff r'\int_{S^2}\diff\Omega\int_{S^2}\diff\Omega'\,\lambda_t(r,\Omega;r',\Omega')\left(\partial_t\rho_t(f)(r,\Omega)\right)\rho_t(g)(r',\Omega')\nonumber\\
		&+\int_0^\infty\diff r\int_0^\infty\diff r'\int_{S^2}\diff\Omega\int_{S^2}\diff\Omega'\,\lambda_t(r,\Omega;r',\Omega')\rho_t(f)(r,\Omega)\left(\partial_t\rho_t(g)(r',\Omega')\right)\nonumber\\
		=&\int_0^\infty\diff r\int_0^\infty\diff r'\int_{S^2}\diff\Omega\int_{S^2}\diff\Omega'\,\left(\partial_t\lambda_t(r,\Omega;r',\Omega')\right)\rho_t(f)(r,\Omega)\rho_t(g)(r',\Omega')\nonumber\\
		&+\int_0^\infty\diff r\int_0^\infty\diff r'\int_{S^2}\diff\Omega\int_{S^2}\diff\Omega'\,\lambda_t(r,\Omega;r',\Omega')\left(\int_r^\infty\diff s\,\left(K_t\rho_t(f)\right)(s,\Omega)\right)\rho_t(g)(r',\Omega')\nonumber\\
		&+\int_0^\infty\diff r\int_0^\infty\diff r'\int_{S^2}\diff\Omega\int_{S^2}\diff\Omega'\,\lambda_t(r,\Omega;r',\Omega')\rho_t(f)(r,\Omega)\left(\int_{r'}^\infty\diff s\,\left(K_t\rho_t(g)\right)(s,\Omega')\right)\nonumber\\
		=&\int_0^\infty\diff r\int_0^\infty\diff r'\int_{S^2}\diff\Omega\int_{S^2}\diff\Omega'\,\left(\partial_t\lambda_t(r,\Omega;r',\Omega')\right)\rho_t(f)(r,\Omega)\rho_t(g)(r',\Omega')\nonumber\\
		&+\int_0^\infty\diff s\int_0^\infty\diff r'\int_{S^2}\diff\Omega\int_{S^2}\diff\Omega'\,\left(\int_0^s\diff r\,\lambda_t(r,\Omega;r',\Omega')\right)\left(K_t\rho_t(f)\right)(s,\Omega)\rho_t(g)(r',\Omega')\nonumber\\
		&+\int_0^\infty\diff r\int_0^\infty\diff s\int_{S^2}\diff\Omega\int_{S^2}\diff\Omega'\,\left(\int_0^s\diff r'\,\lambda_t(r,\Omega;r',\Omega')\right)\rho_t(f)(r,\Omega)\left(K_t\rho_t(g)\right)(s,\Omega')\nonumber\\
		=&\int_0^\infty\diff r\int_0^\infty\diff r'\int_{S^2}\diff\Omega\int_{S^2}\diff\Omega'\,\left(\partial_t\lambda_t(r,\Omega;r',\Omega')\right)\rho_t(f)(r,\Omega)\rho_t(g)(r',\Omega')\nonumber\\
		&+\int_0^\infty\diff r\int_0^\infty\diff r'\int_{S^2}\diff\Omega\int_{S^2}\diff\Omega'\left(\int_0^r\diff s\,\lambda_t(s,\Omega;r',\Omega')\right)\left(K_t\rho_t(f)\right)(r,\Omega)\rho_t(g)(r',\Omega')\nonumber\\
		&+\int_0^\infty\diff r\int_0^\infty\diff r'\int_{S^2}\diff\Omega\int_{S^2}\diff\Omega'\left(\int_0^{r'}\diff s\,\lambda_t(r,\Omega;s,\Omega')\right)\rho_t(f)(r,\Omega)\left(K_t\rho_t(g)\right)(r',\Omega')\nonumber
		\end{align}
		\begin{align}
		=&\int_0^\infty\diff r\int_0^\infty\diff r'\int_{S^2}\diff\Omega\int_{S^2}\diff\Omega'\left(\partial_t\lambda_t(r,\Omega;r',\Omega')\right)\rho_t(f)(r,\Omega)\rho_t(g)(r',\Omega')\nonumber\\
		&+\int_0^\infty\diff r\int_0^\infty\diff r'\int_{S^2}\diff\Omega\int_{S^2}\diff\Omega'\left(K_t\int_0^r\diff s\,\lambda_t(s,\Omega;r',\Omega')\right)\rho_t(f)(r,\Omega)\rho_t(g)(r',\Omega')\nonumber\\
		&+\int_0^\infty\diff r\int_0^\infty\diff r'\int_{S^2}\diff\Omega\int_{S^2}\diff\Omega'\left(K_t'\int_0^{r'}\diff s\,\lambda_t(r,\Omega;s,\Omega')\right)\rho_t(f)(r,\Omega)\rho_t(g)(r',\Omega').
	\end{align}
	This relation holds for all test functions $f,g\in\mathcal{D}(M_t)$, from Ref.\ \cite{hormanderRemarkCharacteristicCauchy1990} we can conclude that the image of $\rho_t$ is dense in an appropriate Sobolev space on $C_t$, which allows us to conclude that
	\begin{equation}
		\partial_t\lambda_t(r,\Omega;r',\Omega')+K_t\int_0^r\diff s\,\lambda_t(s,\Omega;r',\Omega')+K_t'\int_0^{r'}\diff s\,\lambda_t(r,\Omega;s,\Omega')=0.
	\end{equation}
	For the case of Hadamard two-point functions, $\lambda_t$ will in general not be an $L_1$ function. That this argument still goes through can be seen in the following derivation.
	\subsection{Equation \eqref{eq:leading_source}}
	Consider on a general spherically symmetric space-time with metric \eqref{eq:metric} a family of null boundary two-point functions 
	\begin{equation}
		\lambda_t=-\frac{1}{\pi}\frac{\delta(\Omega,\Omega')}{(r-r'-\textup{i}0^+)^2},
	\end{equation}
	on $C_t$ defining Hadamard bulk two-point functions $\Lambda_t$ on $M_t$ via
	\begin{equation}
		\Lambda_t(f,g)=-\lim_{\varepsilon\downarrow 0}\frac{1}{\pi}\int_0^\infty \diff r\int_0^\infty\diff r'\int_{S^2}\diff\Omega\,\frac{\rho_t(f)(r,\Omega)\rho_t(g)(r',\Omega)}{(r-r'-\textup{i}\varepsilon)^2}.
	\end{equation}
	We first note that
	\begin{align}
	&\dv{t}\int_0^\infty \diff r\int_0^\infty\diff r'\int_{S^2}\diff\Omega\,\frac{\rho_t(f)(r,\Omega)\rho_t(g)(r',\Omega)}{(r-r'-\textup{i}\varepsilon)^2}\nonumber\\
	=&\int_0^\infty \diff r\int_0^\infty\diff r'\int_{S^2}\diff\Omega\,\frac{\dot\rho_t(f)(r,\Omega)\rho_t(g)(r',\Omega)}{(r-r'-\textup{i}\varepsilon)^2}\nonumber\\
	&+\int_0^\infty \diff r\int_0^\infty\diff r'\int_{S^2}\diff\Omega\,\frac{\rho_t(f)(r,\Omega)\dot\rho_t(g)(r',\Omega)}{(r-r'-\textup{i}\varepsilon)^2}\nonumber\\
	=&\int_0^\infty \diff r\int_0^\infty\diff r'\int_{S^2}\diff\Omega\int_{S^2}\diff\Omega'\,\rho_t(f)(r,\Omega)\rho_t(g)(r',\Omega')K_t\int_0^r\diff s\,\frac{\delta_{S^2}(\Omega,\Omega')}{(s-r'-\textup{i}\varepsilon)^2}\nonumber\\
	&+\int_0^\infty \diff r\int_0^\infty\diff r'\int_{S^2}\diff\Omega\int_{S^2}\diff\Omega'\,\rho_t(f)(r,\Omega)\rho_t(g)(r',\Omega')K_t'\int_0^{r'}\diff s\,\frac{\delta_{S^2}(\Omega,\Omega')}{(r-s-\textup{i}\varepsilon)^2}\nonumber\\
	=&-\int_0^\infty \diff r\int_0^\infty\diff r'\int_{S^2}\diff\Omega\int_{S^2}\diff\Omega'\,\rho_t(f)(r,\Omega)\rho_t(g)(r',\Omega')K_t\frac{\delta_{S^2}(\Omega,\Omega')}{(r-r'-\textup{i}\varepsilon)}\nonumber\\
	&-\int_0^\infty \diff r\int_0^\infty\diff r'\int_{S^2}\diff\Omega\int_{S^2}\diff\Omega'\,\rho_t(f)(r,\Omega)\rho_t(g)(r',\Omega')K_t\frac{\delta_{S^2}(\Omega,\Omega')}{(r'+\textup{i}\varepsilon)}\nonumber\\
	&+\int_0^\infty \diff r\int_0^\infty\diff r'\int_{S^2}\diff\Omega\int_{S^2}\diff\Omega'\,\rho_t(f)(r,\Omega)\rho_t(g)(r',\Omega')K_t'\frac{\delta_{S^2}(\Omega,\Omega')}{(r-r'-\textup{i}\varepsilon)}\nonumber\\
	&-\int_0^\infty \diff r\int_0^\infty\diff r'\int_{S^2}\diff\Omega\int_{S^2}\diff\Omega'\,\rho_t(f)(r,\Omega)\rho_t(g)(r',\Omega')K_t'\frac{\delta_{S^2}(\Omega,\Omega')}{(r-\textup{i}\varepsilon)}\nonumber
	\end{align}
	\begin{align}
	=&-\frac{1}{2}\int_0^\infty \diff r\int_0^\infty\diff r'\int_{S^2}\diff\Omega\,\rho_t(f)(r,\Omega)\rho_t(g)(r',\Omega)\frac{\exp(2\beta(t,r))\tilde{V}(t,r)-\exp(2\beta(t,r'))\tilde{V}(t,r')}{(r-r'-\textup{i}\varepsilon)}\nonumber\\
	&\begin{aligned}+\frac{1}{2}\int_0^\infty \diff r\int_0^\infty\diff r'\int_{S^2}\diff\Omega\,\rho_t(f)(r,\Omega)\rho_t(g)(r',\Omega)\Big(&2\frac{\exp(2\alpha(t,r))-\exp(2\alpha(t,r'))}{(r-r'-\textup{i}\varepsilon)^3}\\
	&-\frac{(\partial_{r'}\exp(2\alpha(t,r'))+\partial_{r}\exp(2\alpha(t,r)))}{(r-r'-\textup{i}\varepsilon)^2}\Big)\end{aligned}\nonumber\\
	&-\frac{1}{2}\int_0^\infty \diff r\int_0^\infty\diff r'\int_{S^2}\diff\Omega\,\rho_t(f)(r,\Omega)\rho_t(g)(r',\Omega)\frac{\exp(2\beta(t,r))\tilde{V}(t,r)}{(r'+\textup{i}\varepsilon)}\nonumber\\
	&-\frac{1}{2}\int_0^\infty \diff r\int_0^\infty\diff r'\int_{S^2}\diff\Omega\,\rho_t(f)(r,\Omega)\rho_t(g)(r',\Omega)\frac{\exp(2\beta(t,r'))\tilde{V}(t,r')}{(r-\textup{i}\varepsilon)}\nonumber\\
	&\begin{aligned}
	-i\varepsilon\frac{1}{2}\int_0^\infty \diff r\int_0^\infty\diff r'\int_{S^2}\diff\Omega\,\frac{\rho_t(f)(r,\Omega)\Delta_\Omega\rho_t(g)(r',\Omega)}{rr'(r-r'-\textup{i}\varepsilon)}\Bigg(\frac{1}{r'+\textup{i}\varepsilon}+\frac{1}{r-\textup{i}\varepsilon}\Bigg)
	\label{eq:ext_derv2}
	\end{aligned}
	\end{align}
	With some care, one can take the limit $\varepsilon\downarrow 0$ and show that this is uniform in $t$ on a sufficiently small interval. Let us illustrate this for the last line of the equation above. We consider the integral
	\begin{align}
	&\varepsilon\int_0^\infty \diff r\int_0^\infty\diff r'\int_{S^2}\diff\Omega\,\frac{\rho_t(f)(r,\Omega)\Delta_\Omega\rho_t(g)(r',\Omega)}{(r'+\textup{i}\varepsilon)rr'(r-r'-\textup{i}\varepsilon)}\nonumber\\
	=&\varepsilon\int_0^\infty \diff r\int_0^\infty\diff r'\int_{S^2}\diff\Omega\,\frac{\rho_t(f)(r,\Omega)\Delta_\Omega\rho_t(g)(r',\Omega)}{(r'+\textup{i}\varepsilon)rr'}\partial_r\ln(r-r'-\textup{i}\varepsilon)\nonumber\\
	=&-\varepsilon\int_0^\infty\diff r'\int_{S^2}\diff\Omega\,\frac{\lim_{r\downarrow 0}\frac{R}{r}G_c(f)(t,r,\Omega)\Delta_\Omega\rho_t(g)(r',\Omega)}{(r'+\textup{i}\varepsilon)r'}\ln(-r'-\textup{i}\varepsilon)\nonumber\\
	&-\varepsilon\int_0^\infty \diff r\int_0^\infty\diff r'\int_{S^2}\diff\Omega\,\frac{\partial_r\frac{R}{r}G_c(f)(t,r,\Omega)\Delta_\Omega\rho_t(g)(r',\Omega)}{(r'+\textup{i}\varepsilon)r'}\ln(r-r'-\textup{i}\varepsilon).
	\end{align} Note that $G_c(f)(t,0,\Omega)$ is independent of $\Omega$. Therefore, we can easily see
	\begin{equation}
	\int_{S^2}\diff\Omega\lim_{r\downarrow 0}\frac{R}{r}G_c(f)(t,r,\Omega)\Delta_\Omega\rho_t(g)(r',\Omega)=0.
	\end{equation}
	Furthermore, both $\partial_r\frac{R}{r}G_c(f)(t,r,\Omega)$ and $\frac{1}{r^2}\Delta_\Omega\frac{R}{r}G_c(g)(t,r,\Omega)$ are continuous function with spatially compact support, which we shall denote by $F\in C^0(M)$ and $G\in C^0(M)$ respectively. Let $r_m$ be such that for $t\in(a,b)$ some interval we have $F(t,r,\Omega)=G(t,r,\Omega)=0$ for $r>r_m$. We can now estimate
	\begin{align}
	&\varepsilon\left\vert\int_0^\infty \diff r\int_0^\infty\diff r'\int_{S^2}\diff\Omega\,\frac{\rho_t(f)(r,\Omega)\Delta_\Omega\rho_t(g)(r',\Omega)}{(r'+\textup{i}\varepsilon)rr'(r-r'-\textup{i}\varepsilon)}\right\vert\nonumber\\
	\leq&\varepsilon\left\vert\int_0^{r_m} \diff r\int_0^{r_m}\diff r'\int_{S^2}\diff\Omega\,\frac{H(t,r,\Omega){r'}^2G(t,r',\Omega)}{(r'+\textup{i}\varepsilon)}\ln(r-r'-\textup{i}\varepsilon)\right\vert\nonumber\\
	\leq&4\pi\varepsilon\sup_{t\in(a,b),\,r,r'<r_m,\Omega\in S^2}\left\vert F(t,r,\Omega)G(t,r,\Omega)\right\vert\int_0^{r_m} \diff r\int_0^{r_m}\diff r'\,r'\left\vert\ln(r-r'-\textup{i}\varepsilon)\right\vert\nonumber\\
	\leq&4\pi\varepsilon\sup_{t\in(a,b),\,r,r'<r_m,\Omega\in S^2}\left\vert F(t,r,\Omega)G(t,r,\Omega)\right\vert\int_0^{r_m} \diff r\int_0^{r_m}\diff r'\,r'\left(\vert\ln\vert r-r'\vert\vert+\vert\ln(\sqrt{r_m^2+\varepsilon^2})\vert+\pi\right).
	\end{align}
	We now see that the integral in question indeed converges to 0 as $\varepsilon$ tends to 0. In particular, we see this convergence is uniform in $t$ on some finite interval. This uniform convergence can be established for the full integral of Eq.\ \eqref{eq:ext_derv2} using similar arguments. This means in particular that
	\begin{multline}
	\dv{t}\lim_{\varepsilon\downarrow 0}\int_0^\infty \diff r\int_0^\infty\diff r'\int_{S^2}\diff\Omega\frac{\rho_t(f)(r,\Omega)\rho_t(g)(r',\Omega)}{(r-r'-i\varepsilon)^2}\\=\lim_{\varepsilon\downarrow 0}\dv{t}\int_0^\infty \diff r\int_0^\infty\diff r'\int_{S^2}\diff\Omega\frac{\rho_t(f)(r,\Omega)\rho_t(g)(r',\Omega)}{(r-r'-i\varepsilon)^2},
	\end{multline}
	and hence we find Eq. \eqref{eq:leading_source}.
	\bibliography{bibliography}{}
	\bibliographystyle{unsrt}
\end{document}